\documentclass
[aps,prc,twocolumn,amsmath,amssymb,floatfix,noeprint]
{revtex4-1}
\usepackage{lmodern}
\usepackage{CJK}                     
\usepackage[dvips]{graphicx}         
\usepackage{mathptmx}                
\usepackage{physics}                 
\usepackage{dcolumn}                 
\usepackage{multirow}                
\usepackage{xcolor}                  %
\usepackage[breaklinks]{hyperref}
\hypersetup{colorlinks=true,    
 linkcolor=blue,citecolor=blue,filecolor=magenta,urlcolor=blue}

\def\ra{\rightarrow}

\begin{document}

\def\bea{\begin{eqnarray}} \def\eea{\end{eqnarray}}
\def\be{\begin{equation}} \def\ee{\end{equation}}
\def\bal#1\eal{\begin{align}#1\end{align}}
\def\bse#1\ese{\begin{subequations}#1\end{subequations}}
\def\rra{\right\rangle} \def\lla{\left\langle}
\def\al{\alpha}
\def\eps{\epsilon}
\def\mev{\;\text{MeV}}
\def\fm3{\;\text{fm}^{-3}}


\title{
Accurate nuclear symmetry energy at finite temperature
within a BHF approach
}

\begin{CJK*}{UTF8}{gbsn}

\author{Jia-Jing Lu (陆家靖)} 
\author{Fan Li (李凡)}
\author{Zeng-Hua Li (李增花)} \email[]{zhli09@fudan.edu.cn}
\author{Chong-Yang Chen (陈重阳)} 

\affiliation{
Institute of Modern Physics,
Key Laboratory of Nuclear Physics and Ion-beam Application (MOE),
Fudan University, Shanghai 200433, P.R.~China}

\author{G. F. Burgio}
\author{H.-J. Schulze}
\affiliation{
INFN Sezione di Catania, Dipartimento di Fisica,
Universit\'a di Catania, Via Santa Sofia 64, 95123 Catania, Italy}

\date{\today}

\begin{abstract}
We compute the free energy of asymmetric nuclear matter
in a Brueckner-Hartree-Fock approach at finite temperature,
paying particular attention to the dependence on isospin asymmetry.
The first- and second-order symmetry energies are determined
as functions of density and temperature
and useful parametrizations are provided.
We find small deviations from the quadratic isospin dependence
and very small corresponding effects on (proto)neutron star structure.
\end{abstract}


\maketitle
\end{CJK*}

\section{Introduction}

The nuclear symmetry energy,
i.e., the energy difference between removing a neutron or a proton
from nuclear matter
\cite{2016PrPNP..91..203B},
is an important topic of experimental and theoretical nuclear (astro)physics,
as it affects a large number of phenomena
in nuclear structure physics
\cite{2012RPPh...75b6301B},
heavy-ion collisions
\cite{2002Sci...298.1592D,2008PhR...464..113L,2012PhRvC..86a5803T},
and astrophysics like neutron star (NS) structure
\cite{2019EPJA...55..117L,2019arXiv190107673T,2014EPJA...50...40L}
or recently NS mergers
\cite{2017PhRvL.119p1101A,2018PhRvL.121p1101A,2019PhRvX...9a1001A,
2017RPPh...80i6901B,2019PrPNP.10903714B}.

At least under the last two scenarios,
the nuclear system is at non-negligible finite temperature
of the order of several tens of MeV.
This requires to consider the free energy as fundamental
thermodynamical quantity.
Therefore in recent years some phenomenological methods,
such as a momentum-dependent effective interaction \cite{2009PhRvC..79d5806M}
and the nuclear energy-density functional theory \cite{2012JPhCS.342a2003F},
were applied to the study of the behavior of the free energy of nuclear matter
as a function of the baryon density.
More recently, microscopic calculations based on the
self-consistent Green's Function method with
nuclear forces derived from chiral effective field theory
were performed \cite{Carbone_2019}.
Moreover,
we have computed the free energy up to large nucleon densities
$\rho\lesssim0.8\fm3$ and temperatures $T\lesssim50\mev$
within the theoretical Brueckner-Hartree-Fock (BHF) method,
and provided convenient parametrizations for practical use.

Under these circumstances, the nuclear free (symmetry) energy depends on
the partial densities $\rho_n$, $\rho_p$, and temperature $T$.
An important feature is the dependence on isospin asymmetry
$\beta\equiv (\rho_n-\rho_p)/(\rho_n+\rho_p)$
for fixed nucleon density $\rho=\rho_n+\rho_p$,
and for cold matter it has been demonstrated that a quadratic dependence
$\sim \beta^2$ is rather accurate
\cite{1991PhRvC..44.1892B,1999PhRvC..60b4605Z}.
However, at finite temperature this approximation becomes less reliable
\cite{2004PhRvC..69f4001Z,2013NuPhA.902...53T,2017NuPhA.961...78T,
2015PhRvC..92a5801W}
and one should seek to go beyond this lowest-order parametrization.

This is the focus of the present article,
where we study in detail the dependence of the finite-temperature free energy
on isospin and provide parametrizations that go beyond the quadratic law.
We will also give a simple application to NS structure
in order to estimate the magnitude of the effect in practical applications.

We consider in this work two microscopic EOSs
that have been derived within the BHF formalism
\cite{1957RSPSA.239..267G,1976PhR....25...83J,1979NuPhA.328....1D,
Baldo1999,2012RPPh...75b6301B}
based on realistic two-nucleon ($NN$) and compatible three-nucleon forces (TBF)
\cite{1997A&A...328..274B,2004PhRvC..69a8801Z,2002NuPhA.706..418Z,
2008PhRvC..77c4316L,2008PhRvC..78b8801L},
namely those employing the Argonne $V_{18}$ \cite{1995PhRvC..51...38W}
or the Bonn~B \cite{1987PhR...149....1M,Machleidt1989}
$NN$ potentials, respectively.
They all feature reasonable properties at (sub)nuclear densities
in agreement with nuclear-structure phenomenology
\cite{2008PhRvC..78b8801L,2012ChPhL..29a2101L,2016ChPhL..33c2101Q,
2013PhRvC..87d5803T},
and are also fully compatible with recent constraints obtained from the
analysis of the GW170817 NS merger event
\cite{2018ApJ...860..139B,2019JPhG...46c4001W,2020EPJA...56...63W},
as well as from NS cooling
\cite{2018MNRAS.475.5010F,2019MNRAS.484.5162W}.

Our paper is organized as follows.
In Sec.~\ref{s:bhf} we briefly review the computation of the free energy in
the finite-temperature BHF approach
and give some details of the fitting procedure.
In Sec.~\ref{s:res} we present the numerical results
for the free energy and some model calculations of hot NS structure.
Conclusions are drawn in Sec.~\ref{s:end}.

\section{Formalism}
\label{s:bhf}


The calculations for hot asymmetric nuclear matter are based on the
Brueckner-Bethe-Goldstone (BBG) theory
\cite{1957RSPSA.239..267G,1976PhR....25...83J,1979NuPhA.328....1D,
Baldo1999,1991PhRvC..44.1892B,1999PhRvC..60b4605Z}
and its extension to finite temperature
\cite{1986NuPhA.453..189L,1999PhRvC..59..682B,1994PhR...242..165B,
2004PhRvC..69f4001Z}.
Here we simply give a brief review for completeness.
The free energy density in 'frozen-correlations' approximation
\cite{1958NucPh...7..459B,1959NucPh..10..181B,1959NucPh..10..509B,
Baldo1999,1986NuPhA.453..189L,1999PhRvC..59..682B,2006A&A...451..213N,
2006PhRvD..74l3001N,2010PhRvC..81b5806L,2011PhRvC..83b5804B,2010A&A...518A..17B}
is
\be
 f = \rho \frac{F}{A} =
 \sum_{i=n,p} \left[ 2\sum_k n_i(k)
 \left( {k^2\over 2m_i} + {1\over 2}U_i(k) \right) - Ts_i \right] \:,
\label{e:fn}
\ee
where
\be
 s_i = - 2\sum_k \Big( n_i(k) \ln n_i(k) + [1-n_i(k)] \ln [1-n_i(k)] \Big)
\label{eq:entr}
\ee
is the entropy density for the component $i$ treated as a free Fermi gas with
spectrum $e_i(k)$.
At finite temperature,
\be
 n_i(k) =
 \left[\exp{\Big(\frac{e_i(k)-\tilde{\mu}_i}{T}\Big)} + 1 \right]^{-1}
\ee
is a Fermi distribution,
where the auxiliary chemical potentials $\tilde{\mu}_{n,p}$ are fixed
by the condition $\rho_i = 2\sum_k n_i(k)$.
The single-particle energy
\bea
 e_1 &=& \frac{k_1^2}{2m_1} + U_1 \:,
\\
 U_1(\rho,x_p) &=& {\rm Re} \sum_2 n_2
 \langle 1 2| K(\rho,x_p;e_1+e_2) | 1 2 \rangle_a
\label{eq:uk}
\eea
is obtained from the interaction matrix $K$,
which satisfies the self-consistent equation
\be
  K(\rho,x_p;E) = V + V \;\text{Re} \sum_{1,2}
 \frac{|12 \rangle (1-n_1)(1-n_2) \langle 1 2|}
 {E - e_1-e_2 +i0} K(\rho,x_p;E) \:.
\label{eq:BG}
\ee
Here $E$ is the starting energy and
$x_p=\rho_p/\rho$ is the proton fraction.
The multi-indices 1,2 denote in general momentum, isospin, and spin.

Two choices for the realistic $NN$ interaction $V$
are adopted in the present calculations \cite{2008PhRvC..78b8801L}:
the Argonne~$V_{18}$ \cite{1995PhRvC..51...38W}
and the Bonn~B (BOB) \cite{1987PhR...149....1M,Machleidt1989} potential.
They are supplemented with microscopic TBF
employing the same meson-exchange parameters as the two-body potentials.
The TBF are reduced to an effective two-body force
and added to the bare potential in the BHF calculation,
see Refs.~\cite{1989PhRvC..40.1040G,2002NuPhA.706..418Z,
2008PhRvC..77c4316L,2008PhRvC..78b8801L} for details.

The knowledge of the free energy
allows to derive all necessary thermodynamical quantities
in a consistent way, namely
one defines the ``true" chemical potentials $\mu_i$,
pressure $p$, and internal energy density $\eps$ as
\bea
 \mu_i &=& \frac{\partial f}{\partial \rho_i} \:,
\\
 p &=& \rho^2 {\partial{(f/\rho)}\over \partial{\rho}}
 = \sum_i \mu_i \rho_i - f \:,
\label{e:eosp}
\\
 \eps &=& f + Ts \:,\quad
 s = -{{\partial f}\over{\partial T}} \:.
\label{e:eose}
\eea

\begin{table}[t]
\caption{
Parameters of the fit for the free energy per nucleon $F/A$,
Eq.~(\ref{e:fitf}),
for symmetric nuclear matter (SNM),
asymmetric ($\beta=0.6$) nuclear matter (ANM),
and pure neutron matter (PNM)
with the V18 and BOB EOSs.}
\medskip
\def\myc#1{\multicolumn{1}{c}{$#1$}}
\renewcommand{\arraystretch}{1.2}
\begin{ruledtabular}
\begin{tabular}{lr|rrrr|rrrrr}
     &     & \myc{a} & \myc{b} & \myc{c} & \multicolumn{1}{c|}{$d$} &
     $\tilde{a}$ & $\tilde{b}$ & $\tilde{c}$ & $\tilde{d}$ & $\tilde{e}$ \\
\hline\multirow{3}{*}
 {V18}& SNM & -54 & 363 & 2.68 & -8 & -149 & 211 &  -58 &  81 & 2.40 \\
      & ANM & -23 & 473 & 2.72 & -3 & -140 & 200 &  -61 &  82 & 2.36 \\
      & PNM &  38 & 668 & 2.78 &  6 &  -91 & 153 &  -26 &  38 & 2.64 \\
\hline\multirow{3}{*}
 {BOB}& SNM & -60 & 495 & 2.69 & -9 & -124 & 203 &  -60 &  80 & 2.38 \\
      & ANM & -21 & 624 & 2.78 & -4 & -119 & 193 &  -59 &  78 & 2.36 \\
      & PNM &  52 & 860 & 2.89 &  4 &  -82 & 149 &  -25 &  36 & 2.67 \\
\end{tabular}
\end{ruledtabular}
\label{t:fit}
\end{table}

\begin{figure}[t]
\vspace{-3mm}\hspace{-1mm}
\centerline{\includegraphics[scale=0.35]{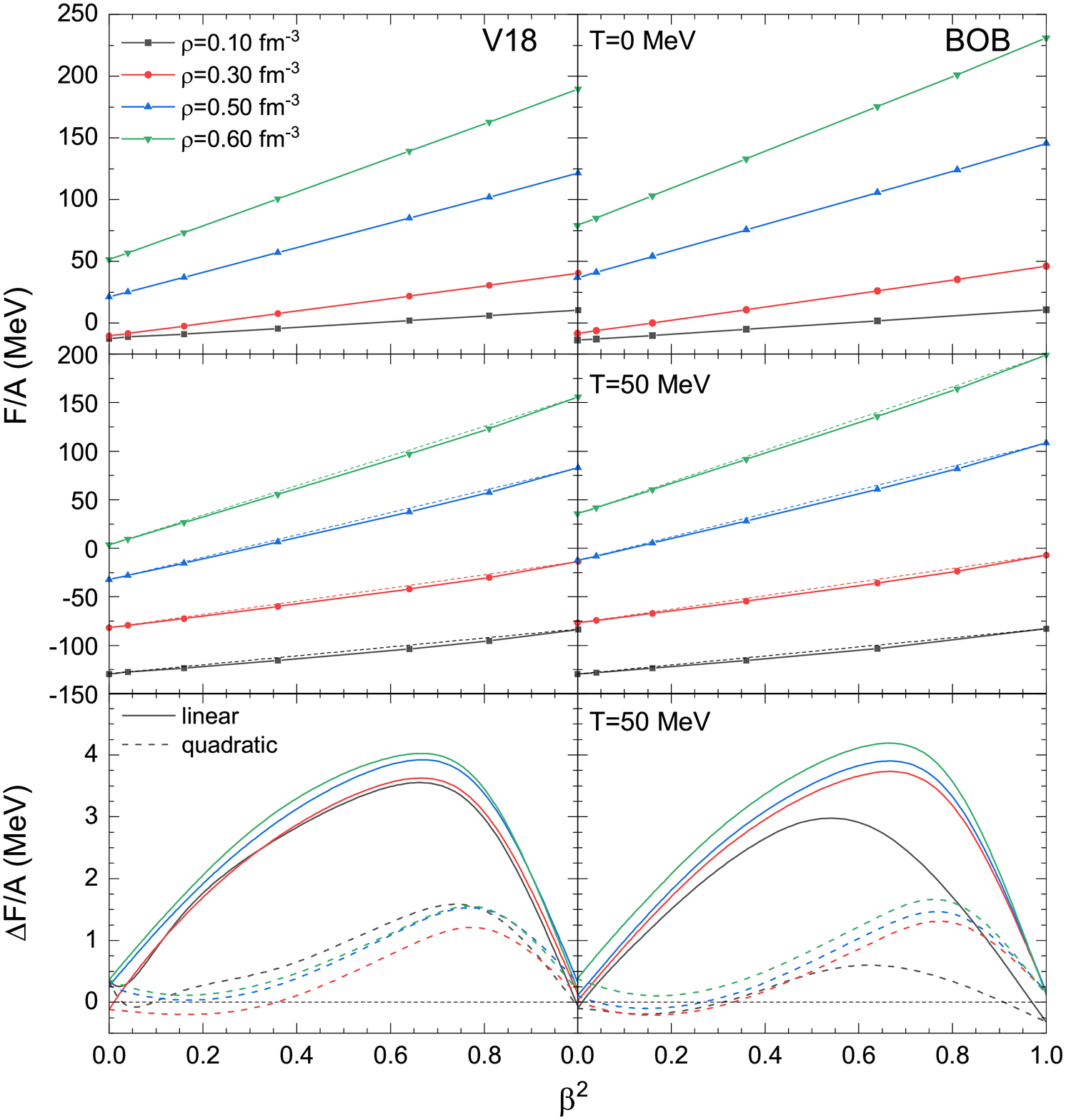}}
\vspace{-10mm}
\caption{
Free energy per nucleon as a function of asymmetry
for different densities at $T=0$ (top panels),
$50\mev$ (middle panels) for the V18 (left panels) or BOB (right panels) EOS.
Dashed lines show the parabolic approximation Eq.~(\ref{e:parab}).
The bottom panels show the deviation between numerical results
and the linear, [Eq.~(\ref{e:parab}), solid curves],
or quadratic, [Eq.~(\ref{e:fsym}), dashed curves],
$\beta^2$ fits.
}
\label{f:fa}
\end{figure}

\begin{figure*}[t]
\vspace{-4mm}
\centerline{\hspace{-2mm}\includegraphics[scale=0.45]{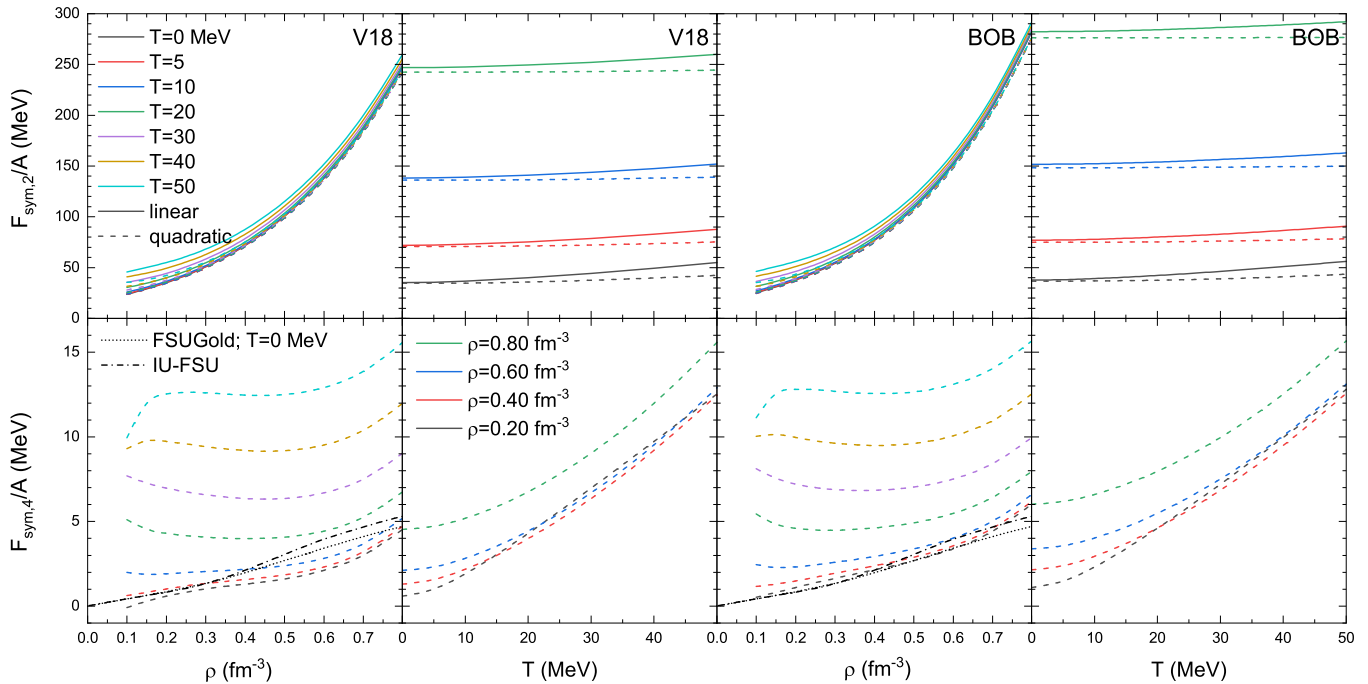}}
\vspace{-5mm}
\caption{
Free symmetry energies per nucleon
$F_\text{sym,2}/A$
(in linear, [Eq.~(\ref{e:parab}), dashed curves],
or quadratic, [Eq.~(\ref{e:fsym}), solid curves],
approximation)
and $F_\text{sym,4}/A$
as functions of nucleon density 
or temperature 
for fixed temperatures/densities, respectively.
For comparison,
the $T=0$ FSUGold and IU-FSU results of Ref.~\cite{2012PhRvC..85b4302C}
are plotted as short dotted and dash-dotted lines respectively in the lower row.
}
\label{f:fsym}
\end{figure*}

For the case of asymmetric nuclear matter,
one might expand the free energy
for fixed total density and temperature
in terms of the asymmetry parameter
$\delta=\beta^2=(1-2x_p)^2$,
\bea
 f(\delta) &\approx& f(0) + \delta f_\text{sym,2} + \delta^2 f_\text{sym,4}
\:.
\label{e:fexp}
\eea
Limiting to the second term, one obtains the symmetry energy
as the difference between pure neutron matter (PNM)
and symmetric nuclear matter (SNM),
\begin{subequations}
\bal
 f_\text{sym,2} &= f(1) - f(0) \:,
\label{e:parab2}
\\
 f_\text{sym,4} &= 0 \:,
\label{e:parab4}
\eal
\label{e:parab}
\end{subequations}
\!\!which is usually a good approximation
at zero temperature
\cite{1991PhRvC..44.1892B,1999PhRvC..60b4605Z,2010A&A...518A..17B},
and also used at finite temperature \cite{2004PhRvC..69f4001Z}.
It has, however, been pointed out
\cite{2013NuPhA.902...53T,2017NuPhA.961...78T,2015PhRvC..92a5801W,
2016PhRvC..93c5806T,2016PhRvC..94b5806N,2019PhRvC..99b5806M,
PhysRevC.96.054311,2019PhRvC.100a5808Z,2018PrPNP..99...29L,
2018PhRvC..97b5801L,2018PhRvC..97e1302W}
that at least the kinetic part of the free energy density
[first term in Eq.~(\ref{e:fn})]
violates the parabolic law, in particular at high temperature.
We therefore extend the expansion to second order
and compute $f_\text{sym,4}$ in the following way:
Inverting the system of equations for $f(0),f(\al),f(1)$,
where $\al$ is an arbitrarily chosen value
(we use $\al=0.6^2$, which corresponds to a typical $x_p=0.2$ in NS matter),
one obtains
\begin{subequations}
\bal
 f_\text{sym,2} &= \frac{\al^2[f(1)-f(0)]-[f(\al)-f(0)]}{\al^2-\al} \:,
\label{e:fsym2}
\\
 f_\text{sym,4} &= \frac{\al[f(1)-f(0)]-[f(\al)-f(0)]}{\al-\al^2} \:,
\label{e:fsym4}
\eal
\label{e:fsym}
\end{subequations}
\!\!in which $f(0),f(\al),f(1)$ depend on total density and temperature.
Following Ref.~\cite{2019PhRvC.100e4335L},
we provide analytical fits for these dependencies
of the numerical results in the required ranges of density
($0.05\fm3 \lesssim \rho \lesssim 1\fm3$)
and temperature ($5\mev \leq T \leq 50\mev$)
in the following functional form for the free energy per nucleon
\bea
{F\over A}(\rho,T) &=&
 a \rho + b \rho^c + d
\nonumber\\&&
 +\, \tilde{a} t^2 \rho
 + \tilde{b} t^2 \ln(\rho)
 + ( \tilde{c} t^2 + \tilde{d} t^{\tilde{e}} )/\rho  \:,
\label{e:fitf}
\eea
where $t=T/(100\mev)$ and $F/A$ and $\rho$ are given in
MeV and $\fm3$, respectively.
The parameters of the fits are listed in Table~\ref{t:fit}
for SNM, asymmetric nuclear matter with $x_p=0.2$ (ANM), and PNM,
for the different EOSs we are using.
The rms deviations of fits and data are better than $0.3\mev$ for all EOSs.

\section{Results}
\label{s:res}

Fig.~\ref{f:fa} shows the free energy per nucleon as a function
of the asymmetry parameter $\delta$
for different densities
and at temperatures $T=0$ (upper row)
and $T=50\mev$ (middle row),
for both EOSs.
The linear approximation Eq.~(\ref{e:fexp},\ref{e:parab})
is indicated by dashed lines in the figure,
and the deviations from the linear
[Eq.~(\ref{e:parab})]
or quadratic
[Eq.~(\ref{e:fsym})] laws
at $T=50\mev$
are indicated in the lower row.
One observes that in general even the linear law provides a very good fit,
even at low density and high temperature,
where the deviations might reach a few percent.
With the quadratic law,
the deviations remain below $2\mev$ over the whole parameter space
$[\rho,T,\beta]$.
In this case
the overall variances
are 0.47 and $0.54\mev$ for the V18 and BOB EOS, respectively.

In order to compare the magnitude of violation
of the linear or quadratic $\beta^2$ laws
with those of other frequently used finite-temperature nuclear EOSs,
we performed the previous analysis also for the SFHo \cite{2013ApJ...774...17S}
and the HShen \cite{1998PThPh.100.1013S,2011ApJS..197...20S} EOS
and report the values of the variance
$\langle \Delta {F/A} \rangle_\text{rms}$
for both the linear and quadratic law
in Table~\ref{t:rms}.
We observe that in all cases the quadratic law is an important improvement
by at least a factor three,
but also the linear law is a very reasonable approximation.

\begin{table}[b]
\caption{
Quality $\langle \Delta {F/A} \rangle_\text{rms}$ (in MeV)
of the linear or quadratic $\beta^2$ laws for the
free energy per nucleon $F/A$
obtained with different EOSs. }
\def\myc#1{\multicolumn{1}{c}{$#1$}}
\def\myc#1{\multicolumn{1}{c}{\text{#1}}}
\renewcommand{\arraystretch}{1.2}
\begin{ruledtabular}
\begin{tabular}{l|dddd}
 EOS           & \myc{V18} & \myc{BOB} & \myc{SFHo} & \myc{Shen} \\
\hline
 linear        & 1.51 & 1.77 & 1.12 & 1.53 \\
 quadratic     & 0.47 & 0.54 & 0.23 & 0.39 \\
\end{tabular}
\end{ruledtabular}
\label{t:rms}
\end{table}

\begin{figure}[t]
\vspace{-3mm}\hspace{-1mm}
\centerline{\includegraphics[scale=0.4]{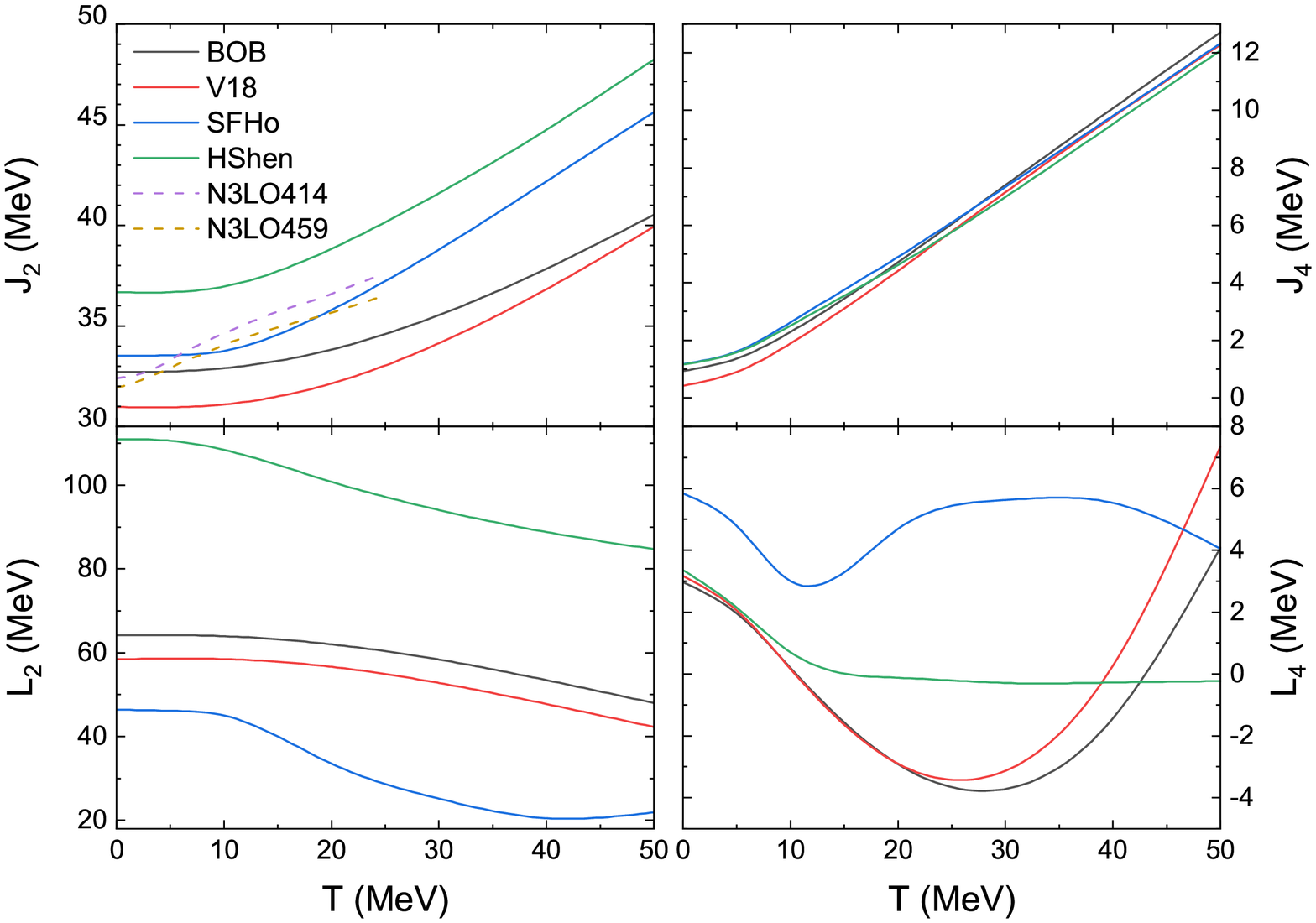}}
\vspace{-8mm}
\caption{
Symmetry energies $J_2,J_4$ (upper panels)
and slope parameters $L_2,L_4$ (lower panels)
at empirical saturation density $\rho_0 = 0.17\fm3$
as a function of temperature
for different EOSs.
The N3LO414 and N3LO450 results of Ref.~\cite{2015PhRvC..92a5801W}
are plotted as dashed curves.
}
\label{f:jlt}
\end{figure}

Fig.~\ref{f:fsym} shows the derived free symmetry energies per nucleon
$\rm F_{sym,2}/A$, Eqs.~(\ref{e:parab2},\ref{e:fsym2}),
and $\rm F_{sym,4}/A$, Eq.~(\ref{e:fsym4}),
as functions of density and temperature.
One notes that the dependence on density is more pronounced for
$\rm F_{sym,2}/A$ than for $\rm F_{sym,4}/A$,
while the opposite is the case for the temperature dependence.
The $\rm F_{sym,2}/A$ results in quadratic approximation
(solid curves in upper row)
are somewhat smaller than in linear approximation (dashed curves)
in order to compensate for the finite $\rm F_{sym,4}/A$,
in particular at finite temperature.
For comparison, the $T=0$ results for $\rm F_{sym,4}/A$
obtained by RMF theory with FSU interactions \cite{2012PhRvC..85b4302C}
are shown as dotted and dash-dotted curves in the lower row.
They are comparable with our BHF results,
especially the BOB model.

The density dependence of the symmetry energies can be expanded around
normal density $\rho_0$ in terms of normal values $J_2,J_4$
and slope parameters $L_2,L_4$:
\bal
 F_\text{sym,2}/\!A(\rho,T) &\approx J_2(T) + L_2(T) x \:,
\\
 F_\text{sym,4}/\!A(\rho,T) &\approx J_4(T) + L_4(T) x \:,
\eal
with $x=(\rho-\rho_0)/3\rho_0$ and
$J_i(T)=J_{\rm sym,i}(\rho_0,T)$,
$L_i(T)=3\partial J_{\rm sym,i}(\rho_0,T)/\partial\rho$.
These quantities are shown in Fig.~\ref{f:jlt}.
The $T=0$ values are
$J_2(0)=31.0(32.7)\mev$ and
$L_2(0)=58.5(64.2)\mev$
for V18(BOB),
which should be confronted with recent constraints
$J_2=31.7\pm2.7\mev$ and
$L_2= 58.7\pm28.1\mev$ \cite{2017RvMP...89a5007O,2019EPJA...55..117L}.
In the same figure we report also the results for the SFHo and Shen EOSs
according to our analysis,
see also Table~\ref{t:rms}.
Reasonable values are obtained in the first case,
but too large ones in the latter.

The second-order symmetry energy $J_4(0)$
is theoretically more controversial compared to the first-order one $J_2(0)$.
Our results are
$J_4(0)=0.41,0.93,1.17,1.17\mev$
for the V18, BOB, SFHo, Shen EOS, respectively.
Within energy density functionals with mean-field approximation,
for example Skyrme-Hartree-Fock and Gogny-Hartree-Fock models,
the values of $J_4$ reported in the literature are around
$1.0\mev$ \cite{PhysRevC.96.054311},
and around $0.66\mev$ within RMF models \cite{2012PhRvC..85b4302C},
while values extracted from Quantum Molecular Dynamics models could be larger
depending on the specific interaction \cite{2016PhRvC..94b5806N}.
From the view point of finite nuclei,
$J_4$ can be related to the second-order symmetry energy $a_{\rm sym,4}(A)$
in a semi-empirical mass formula,
in which the latter can be inferred from the double difference of
``experimental" symmetry energies by analyzing the binding energies
of a large number of measured nuclei
\cite{2017PhLB..773...62W,2014PhRvC..90f4303J}.
In this case, the estimates are $J_4=20.0\pm4.6\mev$ \cite{2017PhLB..773...62W}
and two possible $J_4=8.5\pm0.5\mev$ or $J_4=3.3\pm0.5\mev$
\cite{2014PhRvC..90f4303J},
which are significantly different and larger than those
deduced from nuclear matter,
which points to a great model dependence
and to the importance of finite-size effects in nuclei.

Regarding the temperature dependence,
from Fig.~\ref{f:jlt} one can see that $J_2(T)$ and $J_4(T)$
are increasing monotonically with temperature for all models,
whereas $L_2(T)$ decreases and $L_4(T)$ exhibits nonmonotonic behavior.
It is notable that the $J_4(T)$ results are nearly universal for all EOSs.
Note that in our approach
the temperature dependence is constrained to be a
linear combination of $T^2$ and $T^{\tilde{e}}$ terms
according to Eq.~(\ref{e:fitf}).
We compare our results with the ones of the
chiral effective field theory calculation \cite{2015PhRvC..92a5801W}.
Considering also the cutoff dependence of the chiral potentials,
we observe that both results are in quantitative agreement
in particular in the low temperature region,
but the latter predicts a more linear temperature dependence.
(At low temperature such behavior is excluded by the condition
of vanishing entropy in the $T\ra0$ limit).
The temperature dependence of the free symmetry energy
is also discussed in Refs.~\cite{2007PhRvC..75a4607X,2014EPJA...50...19A},
where an isospin- and momentum-dependent interaction
constrained by heavy-ion collisions and the Skyrme SLy4 parameters
have been employed, respectively.
Those investigations shows very similar behavior and numerical magnitudes
to the present calculations about the free symmetry energy.

\begin{figure}[t]
\vspace{-7mm}\hspace{-0mm}
\centerline{\includegraphics[scale=0.35]{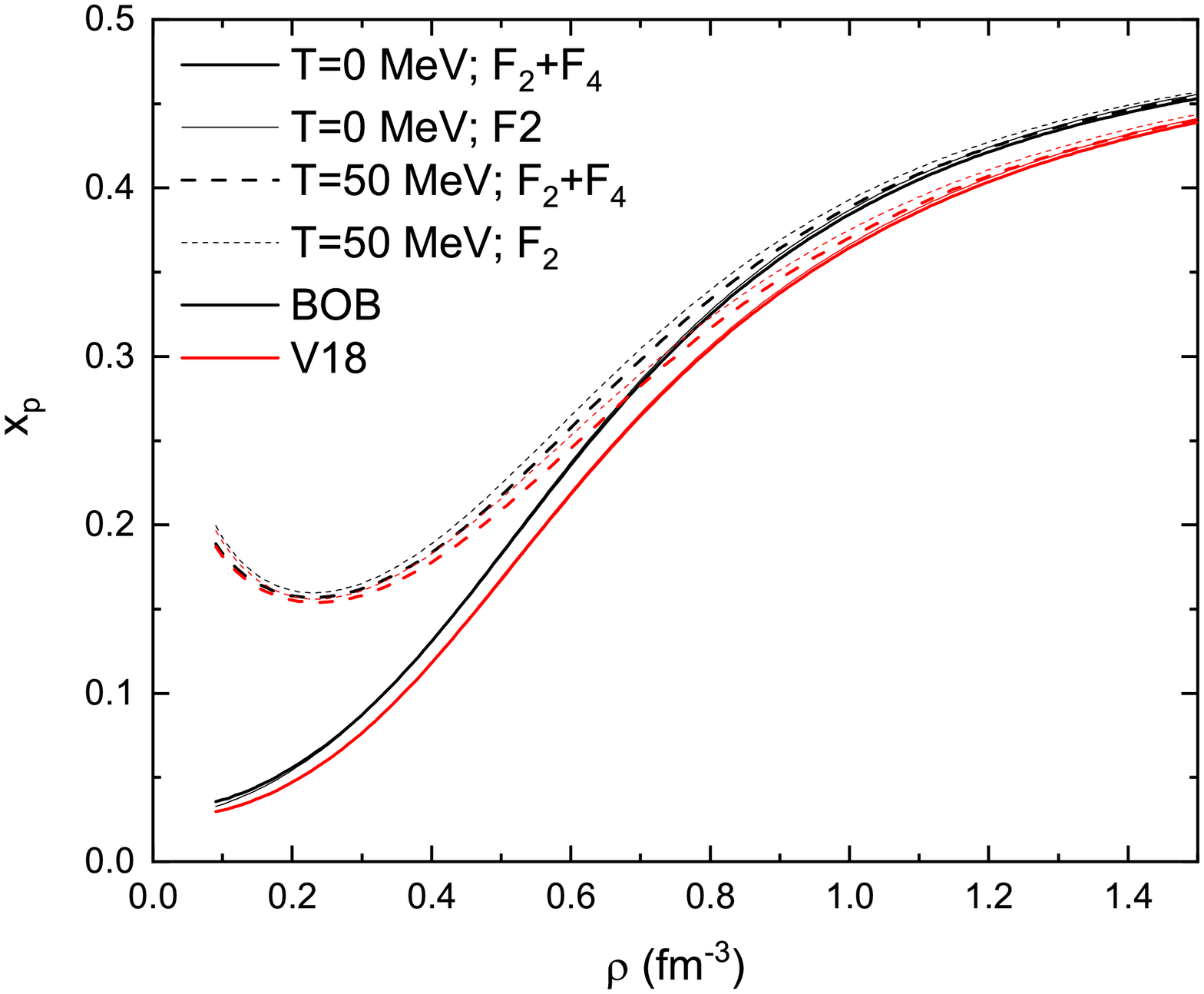}}
\vskip-10mm
\centerline{\includegraphics[scale=0.35]{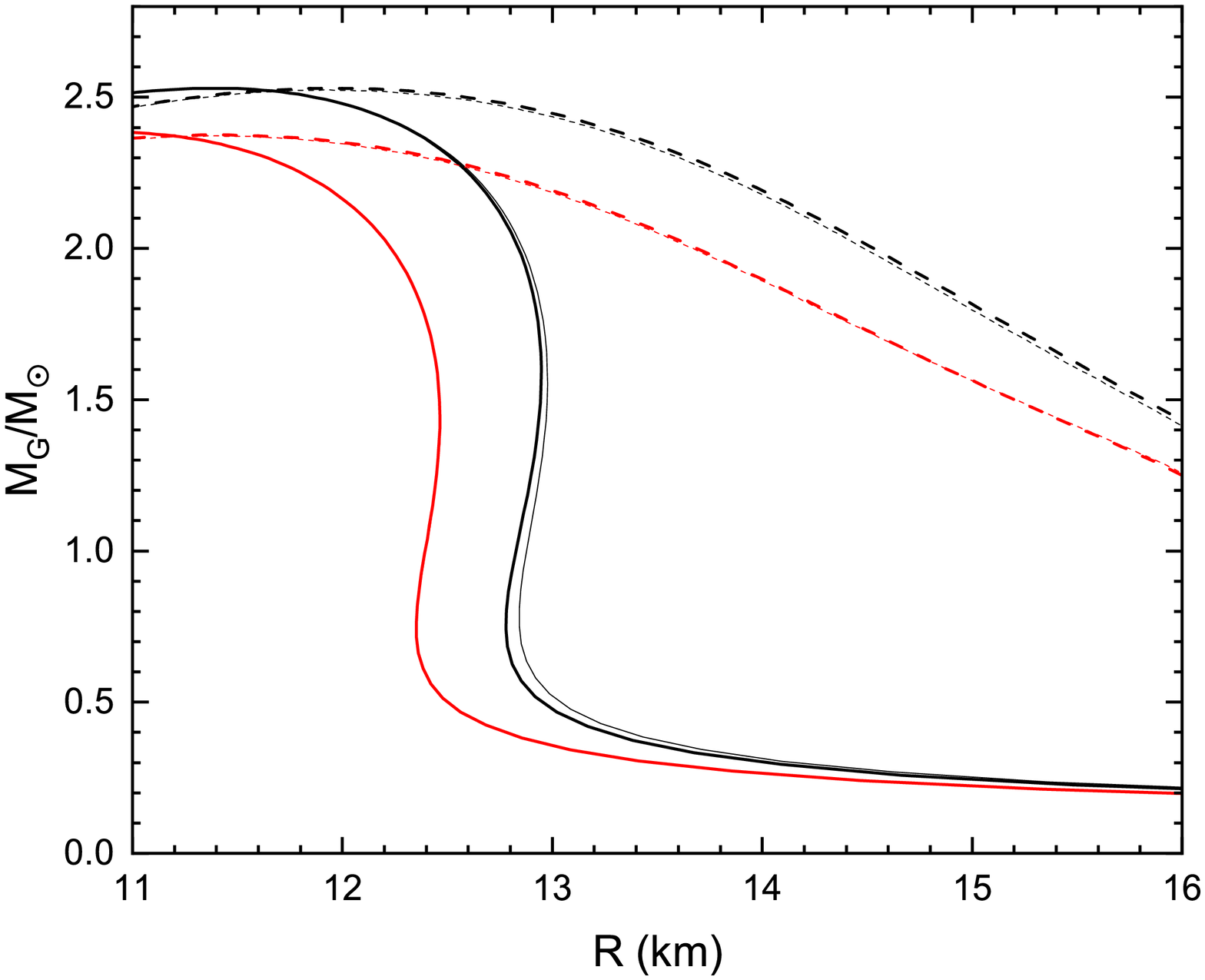}}
\vspace{-7mm}
\caption{
Proton fraction of betastable matter
(upper plot) and
NS mass-radius relation
(lower plot)
at $T=0$ (solid curves)
and $50\mev$ (dashed curves),
employing linear (thin curves)
or quadratic (thick curves)
$\beta^2$ fits, Eqs.~(\ref{e:parab}) or (\ref{e:fsym}).
}
\label{f:ns}
\end{figure}

In order to assess the relevance of the previous results
to practical applications,
we perform some model calculations of NS structure
employing the different approximations for the symmetry energy.
Fig.~\ref{f:ns} shows the proton fractions
of $\beta$-stable and charge-neutral nuclear matter
in the upper panel
and the mass-radius relations of NSs in the lower panel
at the temperatures $T=0$ and $T=50\mev$.
Results using the linear
[Eq.~(\ref{e:parab}), thin curves]
or the quadratic
[Eq.~(\ref{e:fsym}), thick curves]
$\delta$ laws are compared
with both BOB and V18 interactions.
One can see that the inclusion of $F_\text{sym,4}$ in the latter case
causes a slight decrease of the proton fraction
in particular at high temperature,
corresponding to a slight reduction of $F/A$
as seen in Fig.~\ref{f:fa}.
The effect on the mass-radius relations is nearly invisible,
even at large finite temperature,
which means that the linear law Eq.~(\ref{e:parab})
is a very good approximation for the determination of the stellar structure.

\section{Summary}
\label{s:end}

We have studied the isospin dependence
of the free symmetry energy of nuclear matter
at zero and finite temperature
within the framework of the
Brueckner-Hartree-Fock approach at finite temperature
with different potentials and compatible nuclear three-body forces.
We have compared our results with phenomenological models,
i.e., SFHo and Shen EOS,
which are widely used in numerical simulations of astrophysical processes.

We have determined the first- and second-order terms
in an expansion with respect to isospin asymmetry
and provided convenient parametrizations for practical applications.
A model study of neutron star structure at finite temperature
demonstrated that the often used parabolic law
is an excellent approximation
and the second-order modifications are very small.

\section*{Acknowledgments}

This work is sponsored by
the National Natural Science Foundation of China under Grant
Nos.~11475045, 11975077
and the China Scholarship Council, No.~201806100066.
We further acknowledge partial support from ``PHAROS,'' COST Action CA16214.

\vfill

\newcommand{\physrep}{Phys. Rep.}
\newcommand{\nphysa}{Nucl. Phys. A}
\newcommand{\aap}{A\&A}
\newcommand{\mnras}{MNRAS}
\newcommand{\apjs}{ApJS}
\bibliographystyle{apsrev4-2}
\bibliography{articles,books}

\begin{thebibliography}{72}%
\makeatletter
\providecommand \@ifxundefined [1]{%
 \@ifx{#1\undefined}
}%
\providecommand \@ifnum [1]{%
 \ifnum #1\expandafter \@firstoftwo
 \else \expandafter \@secondoftwo
 \fi
}%
\providecommand \@ifx [1]{%
 \ifx #1\expandafter \@firstoftwo
 \else \expandafter \@secondoftwo
 \fi
}%
\providecommand \natexlab [1]{#1}%
\providecommand \enquote  [1]{``#1''}%
\providecommand \bibnamefont  [1]{#1}%
\providecommand \bibfnamefont [1]{#1}%
\providecommand \citenamefont [1]{#1}%
\providecommand \href@noop [0]{\@secondoftwo}%
\providecommand \href [0]{\begingroup \@sanitize@url \@href}%
\providecommand \@href[1]{\@@startlink{#1}\@@href}%
\providecommand \@@href[1]{\endgroup#1\@@endlink}%
\providecommand \@sanitize@url [0]{\catcode `\\12\catcode `\$12\catcode
  `\&12\catcode `\#12\catcode `\^12\catcode `\_12\catcode `\%12\relax}%
\providecommand \@@startlink[1]{}%
\providecommand \@@endlink[0]{}%
\providecommand \url  [0]{\begingroup\@sanitize@url \@url }%
\providecommand \@url [1]{\endgroup\@href {#1}{\urlprefix }}%
\providecommand \urlprefix  [0]{URL }%
\providecommand \Eprint [0]{\href }%
\providecommand \doibase [0]{https://doi.org/}%
\providecommand \selectlanguage [0]{\@gobble}%
\providecommand \bibinfo  [0]{\@secondoftwo}%
\providecommand \bibfield  [0]{\@secondoftwo}%
\providecommand \translation [1]{[#1]}%
\providecommand \BibitemOpen [0]{}%
\providecommand \bibitemStop [0]{}%
\providecommand \bibitemNoStop [0]{.\EOS\space}%
\providecommand \EOS [0]{\spacefactor3000\relax}%
\providecommand \BibitemShut  [1]{\csname bibitem#1\endcsname}%
\let\auto@bib@innerbib\@empty
\bibitem [{\citenamefont {{Baldo}}\ and\ \citenamefont
  {{Burgio}}(2016)}]{2016PrPNP..91..203B}%
  \BibitemOpen
  \bibfield  {author} {\bibinfo {author} {\bibfnamefont {M.}~\bibnamefont
  {{Baldo}}}\ and\ \bibinfo {author} {\bibfnamefont {G.~F.}\ \bibnamefont
  {{Burgio}}},\ }\href {https://doi.org/10.1016/j.ppnp.2016.06.006} {\bibfield
  {journal} {\bibinfo  {journal} {Progress in Particle and Nuclear Physics}\
  }\textbf {\bibinfo {volume} {91}},\ \bibinfo {pages} {203} (\bibinfo {year}
  {2016})},\ \Eprint {https://arxiv.org/abs/1606.08838} {arXiv:1606.08838
  [nucl-th]} \BibitemShut {NoStop}%
\bibitem [{\citenamefont {{Baldo}}\ and\ \citenamefont
  {{Burgio}}(2012)}]{2012RPPh...75b6301B}%
  \BibitemOpen
  \bibfield  {author} {\bibinfo {author} {\bibfnamefont {M.}~\bibnamefont
  {{Baldo}}}\ and\ \bibinfo {author} {\bibfnamefont {G.~F.}\ \bibnamefont
  {{Burgio}}},\ }\href {https://doi.org/10.1088/0034-4885/75/2/026301}
  {\bibfield  {journal} {\bibinfo  {journal} {Reports on Progress in Physics}\
  }\textbf {\bibinfo {volume} {75}},\ \bibinfo {eid} {026301} (\bibinfo {year}
  {2012})},\ \Eprint {https://arxiv.org/abs/1102.1364} {arXiv:1102.1364
  [nucl-th]} \BibitemShut {NoStop}%
\bibitem [{\citenamefont {{Danielewicz}}\ \emph {et~al.}(2002)\citenamefont
  {{Danielewicz}}, \citenamefont {{Lacey}},\ and\ \citenamefont
  {{Lynch}}}]{2002Sci...298.1592D}%
  \BibitemOpen
  \bibfield  {author} {\bibinfo {author} {\bibfnamefont {P.}~\bibnamefont
  {{Danielewicz}}}, \bibinfo {author} {\bibfnamefont {R.}~\bibnamefont
  {{Lacey}}},\ and\ \bibinfo {author} {\bibfnamefont {W.~G.}\ \bibnamefont
  {{Lynch}}},\ }\href {https://doi.org/10.1126/science.1078070} {\bibfield
  {journal} {\bibinfo  {journal} {Science}\ }\textbf {\bibinfo {volume}
  {298}},\ \bibinfo {pages} {1592} (\bibinfo {year} {2002})},\ \Eprint
  {https://arxiv.org/abs/nucl-th/0208016} {arXiv:nucl-th/0208016 [nucl-th]}
  \BibitemShut {NoStop}%
\bibitem [{\citenamefont {{Li}}\ \emph
  {et~al.}(2008{\natexlab{a}})\citenamefont {{Li}}, \citenamefont {{Chen}},\
  and\ \citenamefont {{Ko}}}]{2008PhR...464..113L}%
  \BibitemOpen
  \bibfield  {author} {\bibinfo {author} {\bibfnamefont {B.-A.}\ \bibnamefont
  {{Li}}}, \bibinfo {author} {\bibfnamefont {L.-W.}\ \bibnamefont {{Chen}}},\
  and\ \bibinfo {author} {\bibfnamefont {C.~M.}\ \bibnamefont {{Ko}}},\ }\href
  {https://doi.org/10.1016/j.physrep.2008.04.005} {\bibfield  {journal}
  {\bibinfo  {journal} {\physrep}\ }\textbf {\bibinfo {volume} {464}},\
  \bibinfo {pages} {113} (\bibinfo {year} {2008}{\natexlab{a}})},\ \Eprint
  {https://arxiv.org/abs/0804.3580} {arXiv:0804.3580 [nucl-th]} \BibitemShut
  {NoStop}%
\bibitem [{\citenamefont {{Tsang}}\ \emph {et~al.}(2012)\citenamefont
  {{Tsang}}, \citenamefont {{Stone}}, \citenamefont {{Camera}}, \citenamefont
  {{Danielewicz}}, \citenamefont {{Gandolfi}}, \citenamefont {{Hebeler}},
  \citenamefont {{Horowitz}}, \citenamefont {{Lee}}, \citenamefont {{Lynch}},
  \citenamefont {{Kohley}}, \citenamefont {{Lemmon}}, \citenamefont
  {{M{\"o}ller}}, \citenamefont {{Murakami}}, \citenamefont {{Riordan}},
  \citenamefont {{Roca-Maza}}, \citenamefont {{Sammarruca}}, \citenamefont
  {{Steiner}}, \citenamefont {{Vida{\~n}a}},\ and\ \citenamefont
  {{Yennello}}}]{2012PhRvC..86a5803T}%
  \BibitemOpen
  \bibfield  {author} {\bibinfo {author} {\bibfnamefont {M.~B.}\ \bibnamefont
  {{Tsang}}}, \bibinfo {author} {\bibfnamefont {J.~R.}\ \bibnamefont
  {{Stone}}}, \bibinfo {author} {\bibfnamefont {F.}~\bibnamefont {{Camera}}},
  \bibinfo {author} {\bibfnamefont {P.}~\bibnamefont {{Danielewicz}}}, \bibinfo
  {author} {\bibfnamefont {S.}~\bibnamefont {{Gandolfi}}}, \bibinfo {author}
  {\bibfnamefont {K.}~\bibnamefont {{Hebeler}}}, \bibinfo {author}
  {\bibfnamefont {C.~J.}\ \bibnamefont {{Horowitz}}}, \bibinfo {author}
  {\bibfnamefont {J.}~\bibnamefont {{Lee}}}, \bibinfo {author} {\bibfnamefont
  {W.~G.}\ \bibnamefont {{Lynch}}}, \bibinfo {author} {\bibfnamefont
  {Z.}~\bibnamefont {{Kohley}}}, \bibinfo {author} {\bibfnamefont
  {R.}~\bibnamefont {{Lemmon}}}, \bibinfo {author} {\bibfnamefont
  {P.}~\bibnamefont {{M{\"o}ller}}}, \bibinfo {author} {\bibfnamefont
  {T.}~\bibnamefont {{Murakami}}}, \bibinfo {author} {\bibfnamefont
  {S.}~\bibnamefont {{Riordan}}}, \bibinfo {author} {\bibfnamefont
  {X.}~\bibnamefont {{Roca-Maza}}}, \bibinfo {author} {\bibfnamefont
  {F.}~\bibnamefont {{Sammarruca}}}, \bibinfo {author} {\bibfnamefont {A.~W.}\
  \bibnamefont {{Steiner}}}, \bibinfo {author} {\bibfnamefont {I.}~\bibnamefont
  {{Vida{\~n}a}}},\ and\ \bibinfo {author} {\bibfnamefont {S.~J.}\ \bibnamefont
  {{Yennello}}},\ }\href {https://doi.org/10.1103/PhysRevC.86.015803}
  {\bibfield  {journal} {\bibinfo  {journal} {\prc}\ }\textbf {\bibinfo
  {volume} {86}},\ \bibinfo {eid} {015803} (\bibinfo {year} {2012})},\ \Eprint
  {https://arxiv.org/abs/1204.0466} {arXiv:1204.0466 [nucl-ex]} \BibitemShut
  {NoStop}%
\bibitem [{\citenamefont {{Li}}\ \emph {et~al.}(2019)\citenamefont {{Li}},
  \citenamefont {{Krastev}}, \citenamefont {{Wen}},\ and\ \citenamefont
  {{Zhang}}}]{2019EPJA...55..117L}%
  \BibitemOpen
  \bibfield  {author} {\bibinfo {author} {\bibfnamefont {B.-A.}\ \bibnamefont
  {{Li}}}, \bibinfo {author} {\bibfnamefont {P.~G.}\ \bibnamefont {{Krastev}}},
  \bibinfo {author} {\bibfnamefont {D.-H.}\ \bibnamefont {{Wen}}},\ and\
  \bibinfo {author} {\bibfnamefont {N.-B.}\ \bibnamefont {{Zhang}}},\ }\href
  {https://doi.org/10.1140/epja/i2019-12780-8} {\bibfield  {journal} {\bibinfo
  {journal} {European Physical Journal A}\ }\textbf {\bibinfo {volume} {55}},\
  \bibinfo {eid} {117} (\bibinfo {year} {2019})},\ \Eprint
  {https://arxiv.org/abs/1905.13175} {arXiv:1905.13175 [nucl-th]} \BibitemShut
  {NoStop}%
\bibitem [{\citenamefont {{Tsang}}\ \emph {et~al.}(2019)\citenamefont
  {{Tsang}}, \citenamefont {{Tsang}}, \citenamefont {{Danielewicz}},
  \citenamefont {{Lynch}},\ and\ \citenamefont
  {{Fattoyev}}}]{2019arXiv190107673T}%
  \BibitemOpen
  \bibfield  {author} {\bibinfo {author} {\bibfnamefont {C.~Y.}\ \bibnamefont
  {{Tsang}}}, \bibinfo {author} {\bibfnamefont {M.~B.}\ \bibnamefont
  {{Tsang}}}, \bibinfo {author} {\bibfnamefont {P.}~\bibnamefont
  {{Danielewicz}}}, \bibinfo {author} {\bibfnamefont {W.~G.}\ \bibnamefont
  {{Lynch}}},\ and\ \bibinfo {author} {\bibfnamefont {F.~J.}\ \bibnamefont
  {{Fattoyev}}},\ }\href@noop {} {\bibfield  {journal} {\bibinfo  {journal}
  {arXiv e-prints}\ ,\ \bibinfo {eid} {arXiv:1901.07673}} (\bibinfo {year}
  {2019})},\ \Eprint {https://arxiv.org/abs/1901.07673} {arXiv:1901.07673
  [nucl-ex]} \BibitemShut {NoStop}%
\bibitem [{\citenamefont {{Lattimer}}\ and\ \citenamefont
  {{Steiner}}(2014)}]{2014EPJA...50...40L}%
  \BibitemOpen
  \bibfield  {author} {\bibinfo {author} {\bibfnamefont {J.~M.}\ \bibnamefont
  {{Lattimer}}}\ and\ \bibinfo {author} {\bibfnamefont {A.~W.}\ \bibnamefont
  {{Steiner}}},\ }\href {https://doi.org/10.1140/epja/i2014-14040-y} {\bibfield
   {journal} {\bibinfo  {journal} {European Physical Journal A}\ }\textbf
  {\bibinfo {volume} {50}},\ \bibinfo {eid} {40} (\bibinfo {year} {2014})},\
  \Eprint {https://arxiv.org/abs/1403.1186} {arXiv:1403.1186 [nucl-th]}
  \BibitemShut {NoStop}%
\bibitem [{\citenamefont {{LIGO Scientific Collaboration}}\ and\ \citenamefont
  {{Virgo Collaboration}}(2017)}]{2017PhRvL.119p1101A}%
  \BibitemOpen
  \bibfield  {author} {\bibinfo {author} {\bibnamefont {{LIGO Scientific
  Collaboration}}}\ and\ \bibinfo {author} {\bibnamefont {{Virgo
  Collaboration}}},\ }\href {https://doi.org/10.1103/PhysRevLett.119.161101}
  {\bibfield  {journal} {\bibinfo  {journal} {\prl}\ }\textbf {\bibinfo
  {volume} {119}},\ \bibinfo {eid} {161101} (\bibinfo {year} {2017})},\ \Eprint
  {https://arxiv.org/abs/1710.05832} {arXiv:1710.05832 [gr-qc]} \BibitemShut
  {NoStop}%
\bibitem [{\citenamefont {{LIGO Scientific Collaboration}}\ and\ \citenamefont
  {{Virgo Collaboration}}(2018)}]{2018PhRvL.121p1101A}%
  \BibitemOpen
  \bibfield  {author} {\bibinfo {author} {\bibnamefont {{LIGO Scientific
  Collaboration}}}\ and\ \bibinfo {author} {\bibnamefont {{Virgo
  Collaboration}}},\ }\href {https://doi.org/10.1103/PhysRevLett.121.161101}
  {\bibfield  {journal} {\bibinfo  {journal} {\prl}\ }\textbf {\bibinfo
  {volume} {121}},\ \bibinfo {eid} {161101} (\bibinfo {year} {2018})},\ \Eprint
  {https://arxiv.org/abs/1805.11581} {arXiv:1805.11581 [gr-qc]} \BibitemShut
  {NoStop}%
\bibitem [{\citenamefont {{LIGO Scientific Collaboration}}\ and\ \citenamefont
  {{Virgo Collaboration}}(2019)}]{2019PhRvX...9a1001A}%
  \BibitemOpen
  \bibfield  {author} {\bibinfo {author} {\bibnamefont {{LIGO Scientific
  Collaboration}}}\ and\ \bibinfo {author} {\bibnamefont {{Virgo
  Collaboration}}},\ }\href {https://doi.org/10.1103/PhysRevX.9.011001}
  {\bibfield  {journal} {\bibinfo  {journal} {Physical Review X}\ }\textbf
  {\bibinfo {volume} {9}},\ \bibinfo {eid} {011001} (\bibinfo {year} {2019})},\
  \Eprint {https://arxiv.org/abs/1805.11579} {arXiv:1805.11579 [gr-qc]}
  \BibitemShut {NoStop}%
\bibitem [{\citenamefont {{Baiotti}}\ and\ \citenamefont
  {{Rezzolla}}(2017)}]{2017RPPh...80i6901B}%
  \BibitemOpen
  \bibfield  {author} {\bibinfo {author} {\bibfnamefont {L.}~\bibnamefont
  {{Baiotti}}}\ and\ \bibinfo {author} {\bibfnamefont {L.}~\bibnamefont
  {{Rezzolla}}},\ }\href {https://doi.org/10.1088/1361-6633/aa67bb} {\bibfield
  {journal} {\bibinfo  {journal} {Reports on Progress in Physics}\ }\textbf
  {\bibinfo {volume} {80}},\ \bibinfo {eid} {096901} (\bibinfo {year}
  {2017})},\ \Eprint {https://arxiv.org/abs/1607.03540} {arXiv:1607.03540
  [gr-qc]} \BibitemShut {NoStop}%
\bibitem [{\citenamefont {{Baiotti}}(2019)}]{2019PrPNP.10903714B}%
  \BibitemOpen
  \bibfield  {author} {\bibinfo {author} {\bibfnamefont {L.}~\bibnamefont
  {{Baiotti}}},\ }\href {https://doi.org/10.1016/j.ppnp.2019.103714} {\bibfield
   {journal} {\bibinfo  {journal} {Progress in Particle and Nuclear Physics}\
  }\textbf {\bibinfo {volume} {109}},\ \bibinfo {eid} {103714} (\bibinfo {year}
  {2019})},\ \Eprint {https://arxiv.org/abs/1907.08534} {arXiv:1907.08534
  [astro-ph.HE]} \BibitemShut {NoStop}%
\bibitem [{\citenamefont {{Moustakidis}}\ and\ \citenamefont
  {{Panos}}(2009)}]{2009PhRvC..79d5806M}%
  \BibitemOpen
  \bibfield  {author} {\bibinfo {author} {\bibfnamefont {C.~C.}\ \bibnamefont
  {{Moustakidis}}}\ and\ \bibinfo {author} {\bibfnamefont {C.~P.}\ \bibnamefont
  {{Panos}}},\ }\href {https://doi.org/10.1103/PhysRevC.79.045806} {\bibfield
  {journal} {\bibinfo  {journal} {\prc}\ }\textbf {\bibinfo {volume} {79}},\
  \bibinfo {eid} {045806} (\bibinfo {year} {2009})},\ \Eprint
  {https://arxiv.org/abs/0805.0353} {arXiv:0805.0353 [nucl-th]} \BibitemShut
  {NoStop}%
\bibitem [{\citenamefont {{Fantina}}\ \emph {et~al.}(2012)\citenamefont
  {{Fantina}}, \citenamefont {{Chamel}}, \citenamefont {{Pearson}},\ and\
  \citenamefont {{Goriely}}}]{2012JPhCS.342a2003F}%
  \BibitemOpen
  \bibfield  {author} {\bibinfo {author} {\bibfnamefont {A.~F.}\ \bibnamefont
  {{Fantina}}}, \bibinfo {author} {\bibfnamefont {N.}~\bibnamefont {{Chamel}}},
  \bibinfo {author} {\bibfnamefont {J.~M.}\ \bibnamefont {{Pearson}}},\ and\
  \bibinfo {author} {\bibfnamefont {S.}~\bibnamefont {{Goriely}}},\ }in\ \href
  {https://doi.org/10.1088/1742-6596/342/1/012003} {\emph {\bibinfo {booktitle}
  {Journal of Physics Conference Series}}},\ \bibinfo {series} {Journal of
  Physics Conference Series}, Vol.\ \bibinfo {volume} {342}\ (\bibinfo {year}
  {2012})\ p.\ \bibinfo {pages} {012003}\BibitemShut {NoStop}%
\bibitem [{\citenamefont {Carbone}\ and\ \citenamefont
  {Schwenk}(2019)}]{Carbone_2019}%
  \BibitemOpen
  \bibfield  {author} {\bibinfo {author} {\bibfnamefont {A.}~\bibnamefont
  {Carbone}}\ and\ \bibinfo {author} {\bibfnamefont {A.}~\bibnamefont
  {Schwenk}},\ }\href {https://doi.org/10.1103/physrevc.100.025805} {\bibfield
  {journal} {\bibinfo  {journal} {\prc}\ }\textbf {\bibinfo {volume} {100}},\
  \bibinfo {eid} {025805} (\bibinfo {year} {2019})},\ \Eprint
  {https://arxiv.org/abs/1904.00924} {arXiv:1904.00924} \BibitemShut {NoStop}%
\bibitem [{\citenamefont {{Bombaci}}\ and\ \citenamefont
  {{Lombardo}}(1991)}]{1991PhRvC..44.1892B}%
  \BibitemOpen
  \bibfield  {author} {\bibinfo {author} {\bibfnamefont {I.}~\bibnamefont
  {{Bombaci}}}\ and\ \bibinfo {author} {\bibfnamefont {U.}~\bibnamefont
  {{Lombardo}}},\ }\href {https://doi.org/10.1103/PhysRevC.44.1892} {\bibfield
  {journal} {\bibinfo  {journal} {\prc}\ }\textbf {\bibinfo {volume} {44}},\
  \bibinfo {pages} {1892} (\bibinfo {year} {1991})}\BibitemShut {NoStop}%
\bibitem [{\citenamefont {{Zuo}}\ \emph {et~al.}(1999)\citenamefont {{Zuo}},
  \citenamefont {{Bombaci}},\ and\ \citenamefont
  {{Lombardo}}}]{1999PhRvC..60b4605Z}%
  \BibitemOpen
  \bibfield  {author} {\bibinfo {author} {\bibfnamefont {W.}~\bibnamefont
  {{Zuo}}}, \bibinfo {author} {\bibfnamefont {I.}~\bibnamefont {{Bombaci}}},\
  and\ \bibinfo {author} {\bibfnamefont {U.}~\bibnamefont {{Lombardo}}},\
  }\href {https://doi.org/10.1103/PhysRevC.60.024605} {\bibfield  {journal}
  {\bibinfo  {journal} {\prc}\ }\textbf {\bibinfo {volume} {60}},\ \bibinfo
  {eid} {024605} (\bibinfo {year} {1999})},\ \Eprint
  {https://arxiv.org/abs/nucl-th/0102035} {arXiv:nucl-th/0102035 [nucl-th]}
  \BibitemShut {NoStop}%
\bibitem [{\citenamefont {{Zuo}}\ \emph {et~al.}(2004)\citenamefont {{Zuo}},
  \citenamefont {{Li}}, \citenamefont {{Li}},\ and\ \citenamefont
  {{Lu}}}]{2004PhRvC..69f4001Z}%
  \BibitemOpen
  \bibfield  {author} {\bibinfo {author} {\bibfnamefont {W.}~\bibnamefont
  {{Zuo}}}, \bibinfo {author} {\bibfnamefont {Z.~H.}\ \bibnamefont {{Li}}},
  \bibinfo {author} {\bibfnamefont {A.}~\bibnamefont {{Li}}},\ and\ \bibinfo
  {author} {\bibfnamefont {G.~C.}\ \bibnamefont {{Lu}}},\ }\href
  {https://doi.org/10.1103/PhysRevC.69.064001} {\bibfield  {journal} {\bibinfo
  {journal} {\prc}\ }\textbf {\bibinfo {volume} {69}},\ \bibinfo {eid} {064001}
  (\bibinfo {year} {2004})},\ \Eprint {https://arxiv.org/abs/nucl-th/0412100}
  {arXiv:nucl-th/0412100 [nucl-th]} \BibitemShut {NoStop}%
\bibitem [{\citenamefont {{Togashi}}\ and\ \citenamefont
  {{Takano}}(2013)}]{2013NuPhA.902...53T}%
  \BibitemOpen
  \bibfield  {author} {\bibinfo {author} {\bibfnamefont {H.}~\bibnamefont
  {{Togashi}}}\ and\ \bibinfo {author} {\bibfnamefont {M.}~\bibnamefont
  {{Takano}}},\ }\href {https://doi.org/10.1016/j.nuclphysa.2013.02.014}
  {\bibfield  {journal} {\bibinfo  {journal} {\nphysa}\ }\textbf {\bibinfo
  {volume} {902}},\ \bibinfo {pages} {53} (\bibinfo {year} {2013})},\ \Eprint
  {https://arxiv.org/abs/1302.4261} {arXiv:1302.4261 [nucl-th]} \BibitemShut
  {NoStop}%
\bibitem [{\citenamefont {{Togashi}}\ \emph {et~al.}(2017)\citenamefont
  {{Togashi}}, \citenamefont {{Nakazato}}, \citenamefont {{Takehara}},
  \citenamefont {{Yamamuro}}, \citenamefont {{Suzuki}},\ and\ \citenamefont
  {{Takano}}}]{2017NuPhA.961...78T}%
  \BibitemOpen
  \bibfield  {author} {\bibinfo {author} {\bibfnamefont {H.}~\bibnamefont
  {{Togashi}}}, \bibinfo {author} {\bibfnamefont {K.}~\bibnamefont
  {{Nakazato}}}, \bibinfo {author} {\bibfnamefont {Y.}~\bibnamefont
  {{Takehara}}}, \bibinfo {author} {\bibfnamefont {S.}~\bibnamefont
  {{Yamamuro}}}, \bibinfo {author} {\bibfnamefont {H.}~\bibnamefont
  {{Suzuki}}},\ and\ \bibinfo {author} {\bibfnamefont {M.}~\bibnamefont
  {{Takano}}},\ }\href {https://doi.org/10.1016/j.nuclphysa.2017.02.010}
  {\bibfield  {journal} {\bibinfo  {journal} {\nphysa}\ }\textbf {\bibinfo
  {volume} {961}},\ \bibinfo {pages} {78} (\bibinfo {year} {2017})},\ \Eprint
  {https://arxiv.org/abs/1702.05324} {arXiv:1702.05324 [nucl-th]} \BibitemShut
  {NoStop}%
\bibitem [{\citenamefont {{Wellenhofer}}\ \emph {et~al.}(2015)\citenamefont
  {{Wellenhofer}}, \citenamefont {{Holt}},\ and\ \citenamefont
  {{Kaiser}}}]{2015PhRvC..92a5801W}%
  \BibitemOpen
  \bibfield  {author} {\bibinfo {author} {\bibfnamefont {C.}~\bibnamefont
  {{Wellenhofer}}}, \bibinfo {author} {\bibfnamefont {J.~W.}\ \bibnamefont
  {{Holt}}},\ and\ \bibinfo {author} {\bibfnamefont {N.}~\bibnamefont
  {{Kaiser}}},\ }\href {https://doi.org/10.1103/PhysRevC.92.015801} {\bibfield
  {journal} {\bibinfo  {journal} {\prc}\ }\textbf {\bibinfo {volume} {92}},\
  \bibinfo {eid} {015801} (\bibinfo {year} {2015})},\ \Eprint
  {https://arxiv.org/abs/1504.00177} {arXiv:1504.00177 [nucl-th]} \BibitemShut
  {NoStop}%
\bibitem [{\citenamefont {{Goldstone}}(1957)}]{1957RSPSA.239..267G}%
  \BibitemOpen
  \bibfield  {author} {\bibinfo {author} {\bibfnamefont {J.}~\bibnamefont
  {{Goldstone}}},\ }\href {https://doi.org/10.1098/rspa.1957.0037} {\bibfield
  {journal} {\bibinfo  {journal} {Proceedings of the Royal Society of London
  Series A}\ }\textbf {\bibinfo {volume} {239}},\ \bibinfo {pages} {267}
  (\bibinfo {year} {1957})}\BibitemShut {NoStop}%
\bibitem [{\citenamefont {{Jeukenne}}\ \emph {et~al.}(1976)\citenamefont
  {{Jeukenne}}, \citenamefont {{Lejeune}},\ and\ \citenamefont
  {{Mahaux}}}]{1976PhR....25...83J}%
  \BibitemOpen
  \bibfield  {author} {\bibinfo {author} {\bibfnamefont {J.~P.}\ \bibnamefont
  {{Jeukenne}}}, \bibinfo {author} {\bibfnamefont {A.}~\bibnamefont
  {{Lejeune}}},\ and\ \bibinfo {author} {\bibfnamefont {C.}~\bibnamefont
  {{Mahaux}}},\ }\href {https://doi.org/10.1016/0370-1573(76)90017-X}
  {\bibfield  {journal} {\bibinfo  {journal} {\physrep}\ }\textbf {\bibinfo
  {volume} {25}},\ \bibinfo {pages} {83} (\bibinfo {year} {1976})}\BibitemShut
  {NoStop}%
\bibitem [{\citenamefont {{Day}}(1979)}]{1979NuPhA.328....1D}%
  \BibitemOpen
  \bibfield  {author} {\bibinfo {author} {\bibfnamefont {B.~D.}\ \bibnamefont
  {{Day}}},\ }\href {https://doi.org/10.1016/0375-9474(79)90208-2} {\bibfield
  {journal} {\bibinfo  {journal} {\nphysa}\ }\textbf {\bibinfo {volume}
  {328}},\ \bibinfo {pages} {1} (\bibinfo {year} {1979})}\BibitemShut {NoStop}%
\bibitem [{\citenamefont {{Baldo}}(1999)}]{Baldo1999}%
  \BibitemOpen
  \bibfield  {author} {\bibinfo {author} {\bibfnamefont {M.}~\bibnamefont
  {{Baldo}}},\ }\bibinfo {title} {The many-body theory of the nuclear equation
  of state},\ in\ \href {https://doi.org/10.1142/9789812817501_0001} {\emph
  {\bibinfo {booktitle} {Nuclear Methods and the Nuclear Equation of State}}},\
  \bibinfo {series} {International Review of Nuclear Physics}, Vol.~\bibinfo
  {volume} {8}\ (\bibinfo  {publisher} {World Scientific, Singapore},\ \bibinfo
  {year} {1999})\BibitemShut {NoStop}%
\bibitem [{\citenamefont {{Baldo}}\ \emph {et~al.}(1997)\citenamefont
  {{Baldo}}, \citenamefont {{Bombaci}},\ and\ \citenamefont
  {{Burgio}}}]{1997A&A...328..274B}%
  \BibitemOpen
  \bibfield  {author} {\bibinfo {author} {\bibfnamefont {M.}~\bibnamefont
  {{Baldo}}}, \bibinfo {author} {\bibfnamefont {I.}~\bibnamefont {{Bombaci}}},\
  and\ \bibinfo {author} {\bibfnamefont {G.~F.}\ \bibnamefont {{Burgio}}},\
  }\href@noop {} {\bibfield  {journal} {\bibinfo  {journal} {\aap}\ }\textbf
  {\bibinfo {volume} {328}},\ \bibinfo {pages} {274} (\bibinfo {year}
  {1997})},\ \Eprint {https://arxiv.org/abs/astro-ph/9707277}
  {arXiv:astro-ph/9707277 [astro-ph]} \BibitemShut {NoStop}%
\bibitem [{\citenamefont {{Zhou}}\ \emph {et~al.}(2004)\citenamefont {{Zhou}},
  \citenamefont {{Burgio}}, \citenamefont {{Lombardo}}, \citenamefont
  {{Schulze}},\ and\ \citenamefont {{Zuo}}}]{2004PhRvC..69a8801Z}%
  \BibitemOpen
  \bibfield  {author} {\bibinfo {author} {\bibfnamefont {X.~R.}\ \bibnamefont
  {{Zhou}}}, \bibinfo {author} {\bibfnamefont {G.~F.}\ \bibnamefont
  {{Burgio}}}, \bibinfo {author} {\bibfnamefont {U.}~\bibnamefont
  {{Lombardo}}}, \bibinfo {author} {\bibfnamefont {H.~J.}\ \bibnamefont
  {{Schulze}}},\ and\ \bibinfo {author} {\bibfnamefont {W.}~\bibnamefont
  {{Zuo}}},\ }\href {https://doi.org/10.1103/PhysRevC.69.018801} {\bibfield
  {journal} {\bibinfo  {journal} {\prc}\ }\textbf {\bibinfo {volume} {69}},\
  \bibinfo {eid} {018801} (\bibinfo {year} {2004})}\BibitemShut {NoStop}%
\bibitem [{\citenamefont {{Zuo}}\ \emph {et~al.}(2002)\citenamefont {{Zuo}},
  \citenamefont {{Lejeune}}, \citenamefont {{Lombardo}},\ and\ \citenamefont
  {{Mathiot}}}]{2002NuPhA.706..418Z}%
  \BibitemOpen
  \bibfield  {author} {\bibinfo {author} {\bibfnamefont {W.}~\bibnamefont
  {{Zuo}}}, \bibinfo {author} {\bibfnamefont {A.}~\bibnamefont {{Lejeune}}},
  \bibinfo {author} {\bibfnamefont {U.}~\bibnamefont {{Lombardo}}},\ and\
  \bibinfo {author} {\bibfnamefont {J.~F.}\ \bibnamefont {{Mathiot}}},\ }\href
  {https://doi.org/10.1016/S0375-9474(02)00750-9} {\bibfield  {journal}
  {\bibinfo  {journal} {\nphysa}\ }\textbf {\bibinfo {volume} {706}},\ \bibinfo
  {pages} {418} (\bibinfo {year} {2002})},\ \Eprint
  {https://arxiv.org/abs/nucl-th/0202076} {arXiv:nucl-th/0202076 [nucl-th]}
  \BibitemShut {NoStop}%
\bibitem [{\citenamefont {{Li}}\ \emph
  {et~al.}(2008{\natexlab{b}})\citenamefont {{Li}}, \citenamefont {{Lombardo}},
  \citenamefont {{Schulze}},\ and\ \citenamefont
  {{Zuo}}}]{2008PhRvC..77c4316L}%
  \BibitemOpen
  \bibfield  {author} {\bibinfo {author} {\bibfnamefont {Z.~H.}\ \bibnamefont
  {{Li}}}, \bibinfo {author} {\bibfnamefont {U.}~\bibnamefont {{Lombardo}}},
  \bibinfo {author} {\bibfnamefont {H.~J.}\ \bibnamefont {{Schulze}}},\ and\
  \bibinfo {author} {\bibfnamefont {W.}~\bibnamefont {{Zuo}}},\ }\href
  {https://doi.org/10.1103/PhysRevC.77.034316} {\bibfield  {journal} {\bibinfo
  {journal} {\prc}\ }\textbf {\bibinfo {volume} {77}},\ \bibinfo {eid} {034316}
  (\bibinfo {year} {2008}{\natexlab{b}})}\BibitemShut {NoStop}%
\bibitem [{\citenamefont {{Li}}\ and\ \citenamefont
  {{Schulze}}(2008)}]{2008PhRvC..78b8801L}%
  \BibitemOpen
  \bibfield  {author} {\bibinfo {author} {\bibfnamefont {Z.~H.}\ \bibnamefont
  {{Li}}}\ and\ \bibinfo {author} {\bibfnamefont {H.~J.}\ \bibnamefont
  {{Schulze}}},\ }\href {https://doi.org/10.1103/PhysRevC.78.028801} {\bibfield
   {journal} {\bibinfo  {journal} {\prc}\ }\textbf {\bibinfo {volume} {78}},\
  \bibinfo {eid} {028801} (\bibinfo {year} {2008})}\BibitemShut {NoStop}%
\bibitem [{\citenamefont {{Wiringa}}\ \emph {et~al.}(1995)\citenamefont
  {{Wiringa}}, \citenamefont {{Stoks}},\ and\ \citenamefont
  {{Schiavilla}}}]{1995PhRvC..51...38W}%
  \BibitemOpen
  \bibfield  {author} {\bibinfo {author} {\bibfnamefont {R.~B.}\ \bibnamefont
  {{Wiringa}}}, \bibinfo {author} {\bibfnamefont {V.~G.~J.}\ \bibnamefont
  {{Stoks}}},\ and\ \bibinfo {author} {\bibfnamefont {R.}~\bibnamefont
  {{Schiavilla}}},\ }\href {https://doi.org/10.1103/PhysRevC.51.38} {\bibfield
  {journal} {\bibinfo  {journal} {\prc}\ }\textbf {\bibinfo {volume} {51}},\
  \bibinfo {pages} {38} (\bibinfo {year} {1995})},\ \Eprint
  {https://arxiv.org/abs/nucl-th/9408016} {arXiv:nucl-th/9408016 [nucl-th]}
  \BibitemShut {NoStop}%
\bibitem [{\citenamefont {{Machleidt}}\ \emph {et~al.}(1987)\citenamefont
  {{Machleidt}}, \citenamefont {{Holinde}},\ and\ \citenamefont
  {{Elster}}}]{1987PhR...149....1M}%
  \BibitemOpen
  \bibfield  {author} {\bibinfo {author} {\bibfnamefont {R.}~\bibnamefont
  {{Machleidt}}}, \bibinfo {author} {\bibfnamefont {K.}~\bibnamefont
  {{Holinde}}},\ and\ \bibinfo {author} {\bibfnamefont {C.}~\bibnamefont
  {{Elster}}},\ }\href {https://doi.org/10.1016/S0370-1573(87)80002-9}
  {\bibfield  {journal} {\bibinfo  {journal} {\physrep}\ }\textbf {\bibinfo
  {volume} {149}},\ \bibinfo {pages} {1} (\bibinfo {year} {1987})}\BibitemShut
  {NoStop}%
\bibitem [{\citenamefont {Machleidt}(1989)}]{Machleidt1989}%
  \BibitemOpen
  \bibfield  {author} {\bibinfo {author} {\bibfnamefont {R.}~\bibnamefont
  {Machleidt}},\ }\bibinfo {title} {The meson theory of nuclear forces and
  nuclear structure},\ in\ \href {https://doi.org/10.1007/978-1-4613-9907-0_2}
  {\emph {\bibinfo {booktitle} {Advances in Nuclear Physics}}},\ \bibinfo
  {editor} {edited by\ \bibinfo {editor} {\bibfnamefont {J.~W.}\ \bibnamefont
  {Negele}}\ and\ \bibinfo {editor} {\bibfnamefont {E.}~\bibnamefont {Vogt}}}\
  (\bibinfo  {publisher} {Springer US},\ \bibinfo {address} {Boston, MA},\
  \bibinfo {year} {1989})\ pp.\ \bibinfo {pages} {189--376}\BibitemShut
  {NoStop}%
\bibitem [{\citenamefont {{Li}}\ \emph {et~al.}(2012)\citenamefont {{Li}},
  \citenamefont {{Zhang}}, \citenamefont {{Schulze}},\ and\ \citenamefont
  {{Zuo}}}]{2012ChPhL..29a2101L}%
  \BibitemOpen
  \bibfield  {author} {\bibinfo {author} {\bibfnamefont {Z.-H.}\ \bibnamefont
  {{Li}}}, \bibinfo {author} {\bibfnamefont {D.-P.}\ \bibnamefont {{Zhang}}},
  \bibinfo {author} {\bibfnamefont {H.-J.}\ \bibnamefont {{Schulze}}},\ and\
  \bibinfo {author} {\bibfnamefont {W.}~\bibnamefont {{Zuo}}},\ }\href
  {https://doi.org/10.1088/0256-307X/29/1/012101} {\bibfield  {journal}
  {\bibinfo  {journal} {Chinese Physics Letters}\ }\textbf {\bibinfo {volume}
  {29}},\ \bibinfo {eid} {012101} (\bibinfo {year} {2012})}\BibitemShut
  {NoStop}%
\bibitem [{\citenamefont {{Qing-Yang}}\ \emph {et~al.}(2016)\citenamefont
  {{Qing-Yang}}, \citenamefont {{Zeng-Hua}},\ and\ \citenamefont
  {{Hans-Josef}}}]{2016ChPhL..33c2101Q}%
  \BibitemOpen
  \bibfield  {author} {\bibinfo {author} {\bibfnamefont {B.}~\bibnamefont
  {{Qing-Yang}}}, \bibinfo {author} {\bibfnamefont {L.}~\bibnamefont
  {{Zeng-Hua}}},\ and\ \bibinfo {author} {\bibfnamefont {S.}~\bibnamefont
  {{Hans-Josef}}},\ }\href {https://doi.org/10.1088/0256-307X/33/3/032101}
  {\bibfield  {journal} {\bibinfo  {journal} {Chinese Physics Letters}\
  }\textbf {\bibinfo {volume} {33}},\ \bibinfo {eid} {032101} (\bibinfo {year}
  {2016})}\BibitemShut {NoStop}%
\bibitem [{\citenamefont {{Taranto}}\ \emph {et~al.}(2013)\citenamefont
  {{Taranto}}, \citenamefont {{Baldo}},\ and\ \citenamefont
  {{Burgio}}}]{2013PhRvC..87d5803T}%
  \BibitemOpen
  \bibfield  {author} {\bibinfo {author} {\bibfnamefont {G.}~\bibnamefont
  {{Taranto}}}, \bibinfo {author} {\bibfnamefont {M.}~\bibnamefont {{Baldo}}},\
  and\ \bibinfo {author} {\bibfnamefont {G.~F.}\ \bibnamefont {{Burgio}}},\
  }\href {https://doi.org/10.1103/PhysRevC.87.045803} {\bibfield  {journal}
  {\bibinfo  {journal} {\prc}\ }\textbf {\bibinfo {volume} {87}},\ \bibinfo
  {eid} {045803} (\bibinfo {year} {2013})},\ \Eprint
  {https://arxiv.org/abs/1302.6882} {arXiv:1302.6882 [nucl-th]} \BibitemShut
  {NoStop}%
\bibitem [{\citenamefont {{Burgio}}\ \emph {et~al.}(2018)\citenamefont
  {{Burgio}}, \citenamefont {{Drago}}, \citenamefont {{Pagliara}},
  \citenamefont {{Schulze}},\ and\ \citenamefont
  {{Wei}}}]{2018ApJ...860..139B}%
  \BibitemOpen
  \bibfield  {author} {\bibinfo {author} {\bibfnamefont {G.~F.}\ \bibnamefont
  {{Burgio}}}, \bibinfo {author} {\bibfnamefont {A.}~\bibnamefont {{Drago}}},
  \bibinfo {author} {\bibfnamefont {G.}~\bibnamefont {{Pagliara}}}, \bibinfo
  {author} {\bibfnamefont {H.~J.}\ \bibnamefont {{Schulze}}},\ and\ \bibinfo
  {author} {\bibfnamefont {J.~B.}\ \bibnamefont {{Wei}}},\ }\href
  {https://doi.org/10.3847/1538-4357/aac6ee} {\bibfield  {journal} {\bibinfo
  {journal} {\apj}\ }\textbf {\bibinfo {volume} {860}},\ \bibinfo {eid} {139}
  (\bibinfo {year} {2018})}\BibitemShut {NoStop}%
\bibitem [{\citenamefont {{Wei}}\ \emph
  {et~al.}(2019{\natexlab{a}})\citenamefont {{Wei}}, \citenamefont {{Figura}},
  \citenamefont {{Burgio}}, \citenamefont {{Chen}},\ and\ \citenamefont
  {{Schulze}}}]{2019JPhG...46c4001W}%
  \BibitemOpen
  \bibfield  {author} {\bibinfo {author} {\bibfnamefont {J.~B.}\ \bibnamefont
  {{Wei}}}, \bibinfo {author} {\bibfnamefont {A.}~\bibnamefont {{Figura}}},
  \bibinfo {author} {\bibfnamefont {G.~F.}\ \bibnamefont {{Burgio}}}, \bibinfo
  {author} {\bibfnamefont {H.}~\bibnamefont {{Chen}}},\ and\ \bibinfo {author}
  {\bibfnamefont {H.~J.}\ \bibnamefont {{Schulze}}},\ }\href
  {https://doi.org/10.1088/1361-6471/aaf95c} {\bibfield  {journal} {\bibinfo
  {journal} {Journal of Physics G Nuclear Physics}\ }\textbf {\bibinfo {volume}
  {46}},\ \bibinfo {pages} {034001} (\bibinfo {year} {2019}{\natexlab{a}})},\
  \Eprint {https://arxiv.org/abs/1809.04315} {arXiv:1809.04315 [astro-ph.HE]}
  \BibitemShut {NoStop}%
\bibitem [{\citenamefont {{Wei}}\ \emph {et~al.}(2020)\citenamefont {{Wei}},
  \citenamefont {{Lu}}, \citenamefont {{Burgio}}, \citenamefont {{Li}},\ and\
  \citenamefont {{Schulze}}}]{2020EPJA...56...63W}%
  \BibitemOpen
  \bibfield  {author} {\bibinfo {author} {\bibfnamefont {J.-B.}\ \bibnamefont
  {{Wei}}}, \bibinfo {author} {\bibfnamefont {J.-J.}\ \bibnamefont {{Lu}}},
  \bibinfo {author} {\bibfnamefont {G.~F.}\ \bibnamefont {{Burgio}}}, \bibinfo
  {author} {\bibfnamefont {Z.-H.}\ \bibnamefont {{Li}}},\ and\ \bibinfo
  {author} {\bibfnamefont {H.~J.}\ \bibnamefont {{Schulze}}},\ }\href
  {https://doi.org/10.1140/epja/s10050-020-00058-3} {\bibfield  {journal}
  {\bibinfo  {journal} {European Physical Journal A}\ }\textbf {\bibinfo
  {volume} {56}},\ \bibinfo {eid} {63} (\bibinfo {year} {2020})},\ \Eprint
  {https://arxiv.org/abs/1907.08761} {arXiv:1907.08761 [nucl-th]} \BibitemShut
  {NoStop}%
\bibitem [{\citenamefont {{Fortin}}\ \emph {et~al.}(2018)\citenamefont
  {{Fortin}}, \citenamefont {{Taranto}}, \citenamefont {{Burgio}},
  \citenamefont {{Haensel}}, \citenamefont {{Schulze}},\ and\ \citenamefont
  {{Zdunik}}}]{2018MNRAS.475.5010F}%
  \BibitemOpen
  \bibfield  {author} {\bibinfo {author} {\bibfnamefont {M.}~\bibnamefont
  {{Fortin}}}, \bibinfo {author} {\bibfnamefont {G.}~\bibnamefont {{Taranto}}},
  \bibinfo {author} {\bibfnamefont {G.~F.}\ \bibnamefont {{Burgio}}}, \bibinfo
  {author} {\bibfnamefont {P.}~\bibnamefont {{Haensel}}}, \bibinfo {author}
  {\bibfnamefont {H.~J.}\ \bibnamefont {{Schulze}}},\ and\ \bibinfo {author}
  {\bibfnamefont {J.~L.}\ \bibnamefont {{Zdunik}}},\ }\href
  {https://doi.org/10.1093/mnras/sty147} {\bibfield  {journal} {\bibinfo
  {journal} {\mnras}\ }\textbf {\bibinfo {volume} {475}},\ \bibinfo {pages}
  {5010} (\bibinfo {year} {2018})},\ \Eprint {https://arxiv.org/abs/1709.04855}
  {arXiv:1709.04855 [astro-ph.HE]} \BibitemShut {NoStop}%
\bibitem [{\citenamefont {{Wei}}\ \emph
  {et~al.}(2019{\natexlab{b}})\citenamefont {{Wei}}, \citenamefont {{Burgio}},\
  and\ \citenamefont {{Schulze}}}]{2019MNRAS.484.5162W}%
  \BibitemOpen
  \bibfield  {author} {\bibinfo {author} {\bibfnamefont {J.~B.}\ \bibnamefont
  {{Wei}}}, \bibinfo {author} {\bibfnamefont {G.~F.}\ \bibnamefont
  {{Burgio}}},\ and\ \bibinfo {author} {\bibfnamefont {H.~J.}\ \bibnamefont
  {{Schulze}}},\ }\href {https://doi.org/10.1093/mnras/stz336} {\bibfield
  {journal} {\bibinfo  {journal} {\mnras}\ }\textbf {\bibinfo {volume} {484}},\
  \bibinfo {pages} {5162} (\bibinfo {year} {2019}{\natexlab{b}})},\ \Eprint
  {https://arxiv.org/abs/1812.07306} {arXiv:1812.07306 [astro-ph.HE]}
  \BibitemShut {NoStop}%
\bibitem [{\citenamefont {{Lejeune}}\ \emph {et~al.}(1986)\citenamefont
  {{Lejeune}}, \citenamefont {{Grange}}, \citenamefont {{Martzolff}},\ and\
  \citenamefont {{Cugnon}}}]{1986NuPhA.453..189L}%
  \BibitemOpen
  \bibfield  {author} {\bibinfo {author} {\bibfnamefont {A.}~\bibnamefont
  {{Lejeune}}}, \bibinfo {author} {\bibfnamefont {P.}~\bibnamefont {{Grange}}},
  \bibinfo {author} {\bibfnamefont {M.}~\bibnamefont {{Martzolff}}},\ and\
  \bibinfo {author} {\bibfnamefont {J.}~\bibnamefont {{Cugnon}}},\ }\href
  {https://doi.org/10.1016/0375-9474(86)90010-2} {\bibfield  {journal}
  {\bibinfo  {journal} {\nphysa}\ }\textbf {\bibinfo {volume} {453}},\ \bibinfo
  {pages} {189} (\bibinfo {year} {1986})}\BibitemShut {NoStop}%
\bibitem [{\citenamefont {{Baldo}}\ and\ \citenamefont
  {{Ferreira}}(1999)}]{1999PhRvC..59..682B}%
  \BibitemOpen
  \bibfield  {author} {\bibinfo {author} {\bibfnamefont {M.}~\bibnamefont
  {{Baldo}}}\ and\ \bibinfo {author} {\bibfnamefont {L.~S.}\ \bibnamefont
  {{Ferreira}}},\ }\href {https://doi.org/10.1103/PhysRevC.59.682} {\bibfield
  {journal} {\bibinfo  {journal} {\prc}\ }\textbf {\bibinfo {volume} {59}},\
  \bibinfo {pages} {682} (\bibinfo {year} {1999})}\BibitemShut {NoStop}%
\bibitem [{\citenamefont {{Bombaci}}\ \emph {et~al.}(1994)\citenamefont
  {{Bombaci}}, \citenamefont {{Kuo}},\ and\ \citenamefont
  {{Lombardo}}}]{1994PhR...242..165B}%
  \BibitemOpen
  \bibfield  {author} {\bibinfo {author} {\bibfnamefont {I.}~\bibnamefont
  {{Bombaci}}}, \bibinfo {author} {\bibfnamefont {T.~T.~S.}\ \bibnamefont
  {{Kuo}}},\ and\ \bibinfo {author} {\bibfnamefont {U.}~\bibnamefont
  {{Lombardo}}},\ }\href {https://doi.org/10.1016/0370-1573(94)90149-X}
  {\bibfield  {journal} {\bibinfo  {journal} {\physrep}\ }\textbf {\bibinfo
  {volume} {242}},\ \bibinfo {pages} {165} (\bibinfo {year}
  {1994})}\BibitemShut {NoStop}%
\bibitem [{\citenamefont {{Bloch}}\ and\ \citenamefont {{De
  Dominicis}}(1958)}]{1958NucPh...7..459B}%
  \BibitemOpen
  \bibfield  {author} {\bibinfo {author} {\bibfnamefont {C.}~\bibnamefont
  {{Bloch}}}\ and\ \bibinfo {author} {\bibfnamefont {C.}~\bibnamefont {{De
  Dominicis}}},\ }\href {https://doi.org/10.1016/0029-5582(58)90285-2}
  {\bibfield  {journal} {\bibinfo  {journal} {Nuclear Physics}\ }\textbf
  {\bibinfo {volume} {7}},\ \bibinfo {pages} {459} (\bibinfo {year}
  {1958})}\BibitemShut {NoStop}%
\bibitem [{\citenamefont {{Bloch}}\ and\ \citenamefont {{De
  Dominicis}}(1959{\natexlab{a}})}]{1959NucPh..10..181B}%
  \BibitemOpen
  \bibfield  {author} {\bibinfo {author} {\bibfnamefont {C.}~\bibnamefont
  {{Bloch}}}\ and\ \bibinfo {author} {\bibfnamefont {C.}~\bibnamefont {{De
  Dominicis}}},\ }\href {https://doi.org/10.1016/0029-5582(59)90203-2}
  {\bibfield  {journal} {\bibinfo  {journal} {Nuclear Physics}\ }\textbf
  {\bibinfo {volume} {10}},\ \bibinfo {pages} {181} (\bibinfo {year}
  {1959}{\natexlab{a}})}\BibitemShut {NoStop}%
\bibitem [{\citenamefont {{Bloch}}\ and\ \citenamefont {{De
  Dominicis}}(1959{\natexlab{b}})}]{1959NucPh..10..509B}%
  \BibitemOpen
  \bibfield  {author} {\bibinfo {author} {\bibfnamefont {C.}~\bibnamefont
  {{Bloch}}}\ and\ \bibinfo {author} {\bibfnamefont {C.}~\bibnamefont {{De
  Dominicis}}},\ }\href {https://doi.org/10.1016/0029-5582(59)90241-X}
  {\bibfield  {journal} {\bibinfo  {journal} {Nuclear Physics}\ }\textbf
  {\bibinfo {volume} {10}},\ \bibinfo {pages} {509} (\bibinfo {year}
  {1959}{\natexlab{b}})}\BibitemShut {NoStop}%
\bibitem [{\citenamefont {{Nicotra}}\ \emph
  {et~al.}(2006{\natexlab{a}})\citenamefont {{Nicotra}}, \citenamefont
  {{Baldo}}, \citenamefont {{Burgio}},\ and\ \citenamefont
  {{Schulze}}}]{2006A&A...451..213N}%
  \BibitemOpen
  \bibfield  {author} {\bibinfo {author} {\bibfnamefont {O.~E.}\ \bibnamefont
  {{Nicotra}}}, \bibinfo {author} {\bibfnamefont {M.}~\bibnamefont {{Baldo}}},
  \bibinfo {author} {\bibfnamefont {G.~F.}\ \bibnamefont {{Burgio}}},\ and\
  \bibinfo {author} {\bibfnamefont {H.~J.}\ \bibnamefont {{Schulze}}},\ }\href
  {https://doi.org/10.1051/0004-6361:20053670} {\bibfield  {journal} {\bibinfo
  {journal} {\aap}\ }\textbf {\bibinfo {volume} {451}},\ \bibinfo {pages} {213}
  (\bibinfo {year} {2006}{\natexlab{a}})},\ \Eprint
  {https://arxiv.org/abs/nucl-th/0506066} {arXiv:nucl-th/0506066 [nucl-th]}
  \BibitemShut {NoStop}%
\bibitem [{\citenamefont {{Nicotra}}\ \emph
  {et~al.}(2006{\natexlab{b}})\citenamefont {{Nicotra}}, \citenamefont
  {{Baldo}}, \citenamefont {{Burgio}},\ and\ \citenamefont
  {{Schulze}}}]{2006PhRvD..74l3001N}%
  \BibitemOpen
  \bibfield  {author} {\bibinfo {author} {\bibfnamefont {O.~E.}\ \bibnamefont
  {{Nicotra}}}, \bibinfo {author} {\bibfnamefont {M.}~\bibnamefont {{Baldo}}},
  \bibinfo {author} {\bibfnamefont {G.~F.}\ \bibnamefont {{Burgio}}},\ and\
  \bibinfo {author} {\bibfnamefont {H.~J.}\ \bibnamefont {{Schulze}}},\ }\href
  {https://doi.org/10.1103/PhysRevD.74.123001} {\bibfield  {journal} {\bibinfo
  {journal} {\prd}\ }\textbf {\bibinfo {volume} {74}},\ \bibinfo {eid} {123001}
  (\bibinfo {year} {2006}{\natexlab{b}})},\ \Eprint
  {https://arxiv.org/abs/astro-ph/0608021} {arXiv:astro-ph/0608021 [astro-ph]}
  \BibitemShut {NoStop}%
\bibitem [{\citenamefont {{Li}}\ \emph {et~al.}(2010)\citenamefont {{Li}},
  \citenamefont {{Zhou}}, \citenamefont {{Burgio}},\ and\ \citenamefont
  {{Schulze}}}]{2010PhRvC..81b5806L}%
  \BibitemOpen
  \bibfield  {author} {\bibinfo {author} {\bibfnamefont {A.}~\bibnamefont
  {{Li}}}, \bibinfo {author} {\bibfnamefont {X.~R.}\ \bibnamefont {{Zhou}}},
  \bibinfo {author} {\bibfnamefont {G.~F.}\ \bibnamefont {{Burgio}}},\ and\
  \bibinfo {author} {\bibfnamefont {H.~J.}\ \bibnamefont {{Schulze}}},\ }\href
  {https://doi.org/10.1103/PhysRevC.81.025806} {\bibfield  {journal} {\bibinfo
  {journal} {\prc}\ }\textbf {\bibinfo {volume} {81}},\ \bibinfo {eid} {025806}
  (\bibinfo {year} {2010})},\ \Eprint {https://arxiv.org/abs/1002.0642}
  {arXiv:1002.0642 [astro-ph.SR]} \BibitemShut {NoStop}%
\bibitem [{\citenamefont {{Burgio}}\ \emph {et~al.}(2011)\citenamefont
  {{Burgio}}, \citenamefont {{Schulze}},\ and\ \citenamefont
  {{Li}}}]{2011PhRvC..83b5804B}%
  \BibitemOpen
  \bibfield  {author} {\bibinfo {author} {\bibfnamefont {G.~F.}\ \bibnamefont
  {{Burgio}}}, \bibinfo {author} {\bibfnamefont {H.~J.}\ \bibnamefont
  {{Schulze}}},\ and\ \bibinfo {author} {\bibfnamefont {A.}~\bibnamefont
  {{Li}}},\ }\href {https://doi.org/10.1103/PhysRevC.83.025804} {\bibfield
  {journal} {\bibinfo  {journal} {\prc}\ }\textbf {\bibinfo {volume} {83}},\
  \bibinfo {eid} {025804} (\bibinfo {year} {2011})},\ \Eprint
  {https://arxiv.org/abs/1101.0726} {arXiv:1101.0726 [astro-ph.SR]}
  \BibitemShut {NoStop}%
\bibitem [{\citenamefont {{Burgio}}\ and\ \citenamefont
  {{Schulze}}(2010)}]{2010A&A...518A..17B}%
  \BibitemOpen
  \bibfield  {author} {\bibinfo {author} {\bibfnamefont {G.~F.}\ \bibnamefont
  {{Burgio}}}\ and\ \bibinfo {author} {\bibfnamefont {H.~J.}\ \bibnamefont
  {{Schulze}}},\ }\href {https://doi.org/10.1051/0004-6361/201014308}
  {\bibfield  {journal} {\bibinfo  {journal} {\aap}\ }\textbf {\bibinfo
  {volume} {518}},\ \bibinfo {eid} {A17} (\bibinfo {year} {2010})},\ \Eprint
  {https://arxiv.org/abs/1002.4497} {arXiv:1002.4497 [astro-ph.SR]}
  \BibitemShut {NoStop}%
\bibitem [{\citenamefont {{Grang{\'e}}}\ \emph {et~al.}(1989)\citenamefont
  {{Grang{\'e}}}, \citenamefont {{Lejeune}}, \citenamefont {{Martzolff}},\ and\
  \citenamefont {{Mathiot}}}]{1989PhRvC..40.1040G}%
  \BibitemOpen
  \bibfield  {author} {\bibinfo {author} {\bibfnamefont {P.}~\bibnamefont
  {{Grang{\'e}}}}, \bibinfo {author} {\bibfnamefont {A.}~\bibnamefont
  {{Lejeune}}}, \bibinfo {author} {\bibfnamefont {M.}~\bibnamefont
  {{Martzolff}}},\ and\ \bibinfo {author} {\bibfnamefont {J.~F.}\ \bibnamefont
  {{Mathiot}}},\ }\href {https://doi.org/10.1103/PhysRevC.40.1040} {\bibfield
  {journal} {\bibinfo  {journal} {\prc}\ }\textbf {\bibinfo {volume} {40}},\
  \bibinfo {pages} {1040} (\bibinfo {year} {1989})}\BibitemShut {NoStop}%
\bibitem [{\citenamefont {{Cai}}\ and\ \citenamefont
  {{Chen}}(2012)}]{2012PhRvC..85b4302C}%
  \BibitemOpen
  \bibfield  {author} {\bibinfo {author} {\bibfnamefont {B.-J.}\ \bibnamefont
  {{Cai}}}\ and\ \bibinfo {author} {\bibfnamefont {L.-W.}\ \bibnamefont
  {{Chen}}},\ }\href {https://doi.org/10.1103/PhysRevC.85.024302} {\bibfield
  {journal} {\bibinfo  {journal} {\prc}\ }\textbf {\bibinfo {volume} {85}},\
  \bibinfo {eid} {024302} (\bibinfo {year} {2012})},\ \Eprint
  {https://arxiv.org/abs/1111.4124} {arXiv:1111.4124 [nucl-th]} \BibitemShut
  {NoStop}%
\bibitem [{\citenamefont {{Tan}}\ \emph {et~al.}(2016)\citenamefont {{Tan}},
  \citenamefont {{Loan}}, \citenamefont {{Khoa}},\ and\ \citenamefont
  {{Margueron}}}]{2016PhRvC..93c5806T}%
  \BibitemOpen
  \bibfield  {author} {\bibinfo {author} {\bibfnamefont {N.~H.}\ \bibnamefont
  {{Tan}}}, \bibinfo {author} {\bibfnamefont {D.~T.}\ \bibnamefont {{Loan}}},
  \bibinfo {author} {\bibfnamefont {D.~T.}\ \bibnamefont {{Khoa}}},\ and\
  \bibinfo {author} {\bibfnamefont {J.}~\bibnamefont {{Margueron}}},\ }\href
  {https://doi.org/10.1103/PhysRevC.93.035806} {\bibfield  {journal} {\bibinfo
  {journal} {\prc}\ }\textbf {\bibinfo {volume} {93}},\ \bibinfo {eid} {035806}
  (\bibinfo {year} {2016})}\BibitemShut {NoStop}%
\bibitem [{\citenamefont {{Nandi}}\ and\ \citenamefont
  {{Schramm}}(2016)}]{2016PhRvC..94b5806N}%
  \BibitemOpen
  \bibfield  {author} {\bibinfo {author} {\bibfnamefont {R.}~\bibnamefont
  {{Nandi}}}\ and\ \bibinfo {author} {\bibfnamefont {S.}~\bibnamefont
  {{Schramm}}},\ }\href {https://doi.org/10.1103/PhysRevC.94.025806} {\bibfield
   {journal} {\bibinfo  {journal} {\prc}\ }\textbf {\bibinfo {volume} {94}},\
  \bibinfo {eid} {025806} (\bibinfo {year} {2016})},\ \Eprint
  {https://arxiv.org/abs/1601.01842} {arXiv:1601.01842 [nucl-th]} \BibitemShut
  {NoStop}%
\bibitem [{\citenamefont {{Margueron}}\ and\ \citenamefont
  {{Gulminelli}}(2019)}]{2019PhRvC..99b5806M}%
  \BibitemOpen
  \bibfield  {author} {\bibinfo {author} {\bibfnamefont {J.}~\bibnamefont
  {{Margueron}}}\ and\ \bibinfo {author} {\bibfnamefont {F.}~\bibnamefont
  {{Gulminelli}}},\ }\href {https://doi.org/10.1103/PhysRevC.99.025806}
  {\bibfield  {journal} {\bibinfo  {journal} {\prc}\ }\textbf {\bibinfo
  {volume} {99}},\ \bibinfo {eid} {025806} (\bibinfo {year} {2019})},\ \Eprint
  {https://arxiv.org/abs/1807.01729} {arXiv:1807.01729 [nucl-th]} \BibitemShut
  {NoStop}%
\bibitem [{\citenamefont {Pu}\ \emph {et~al.}(2017)\citenamefont {Pu},
  \citenamefont {Zhang},\ and\ \citenamefont {Chen}}]{PhysRevC.96.054311}%
  \BibitemOpen
  \bibfield  {author} {\bibinfo {author} {\bibfnamefont {J.}~\bibnamefont
  {Pu}}, \bibinfo {author} {\bibfnamefont {Z.}~\bibnamefont {Zhang}},\ and\
  \bibinfo {author} {\bibfnamefont {L.-W.}\ \bibnamefont {Chen}},\ }\href
  {https://doi.org/10.1103/PhysRevC.96.054311} {\bibfield  {journal} {\bibinfo
  {journal} {Phys. Rev. C}\ }\textbf {\bibinfo {volume} {96}},\ \bibinfo
  {pages} {054311} (\bibinfo {year} {2017})}\BibitemShut {NoStop}%
\bibitem [{\citenamefont {{Zabari}}\ \emph {et~al.}(2019)\citenamefont
  {{Zabari}}, \citenamefont {{Kubis}},\ and\ \citenamefont
  {{W{\'o}jcik}}}]{2019PhRvC.100a5808Z}%
  \BibitemOpen
  \bibfield  {author} {\bibinfo {author} {\bibfnamefont {N.}~\bibnamefont
  {{Zabari}}}, \bibinfo {author} {\bibfnamefont {S.}~\bibnamefont {{Kubis}}},\
  and\ \bibinfo {author} {\bibfnamefont {W.}~\bibnamefont {{W{\'o}jcik}}},\
  }\href {https://doi.org/10.1103/PhysRevC.100.015808} {\bibfield  {journal}
  {\bibinfo  {journal} {\prc}\ }\textbf {\bibinfo {volume} {100}},\ \bibinfo
  {eid} {015808} (\bibinfo {year} {2019})},\ \Eprint
  {https://arxiv.org/abs/1908.00476} {arXiv:1908.00476 [nucl-th]} \BibitemShut
  {NoStop}%
\bibitem [{\citenamefont {{Li}}\ \emph {et~al.}(2018)\citenamefont {{Li}},
  \citenamefont {{Cai}}, \citenamefont {{Chen}},\ and\ \citenamefont
  {{Xu}}}]{2018PrPNP..99...29L}%
  \BibitemOpen
  \bibfield  {author} {\bibinfo {author} {\bibfnamefont {B.-A.}\ \bibnamefont
  {{Li}}}, \bibinfo {author} {\bibfnamefont {B.-J.}\ \bibnamefont {{Cai}}},
  \bibinfo {author} {\bibfnamefont {L.-W.}\ \bibnamefont {{Chen}}},\ and\
  \bibinfo {author} {\bibfnamefont {J.}~\bibnamefont {{Xu}}},\ }\href
  {https://doi.org/10.1016/j.ppnp.2018.01.001} {\bibfield  {journal} {\bibinfo
  {journal} {Progress in Particle and Nuclear Physics}\ }\textbf {\bibinfo
  {volume} {99}},\ \bibinfo {pages} {29} (\bibinfo {year} {2018})},\ \Eprint
  {https://arxiv.org/abs/1801.01213} {arXiv:1801.01213 [nucl-th]} \BibitemShut
  {NoStop}%
\bibitem [{\citenamefont {{Liu}}\ \emph {et~al.}(2018)\citenamefont {{Liu}},
  \citenamefont {{Qian}}, \citenamefont {{Xing}}, \citenamefont {{Niu}},\ and\
  \citenamefont {{Sun}}}]{2018PhRvC..97b5801L}%
  \BibitemOpen
  \bibfield  {author} {\bibinfo {author} {\bibfnamefont {Z.~W.}\ \bibnamefont
  {{Liu}}}, \bibinfo {author} {\bibfnamefont {Z.}~\bibnamefont {{Qian}}},
  \bibinfo {author} {\bibfnamefont {R.~Y.}\ \bibnamefont {{Xing}}}, \bibinfo
  {author} {\bibfnamefont {J.~R.}\ \bibnamefont {{Niu}}},\ and\ \bibinfo
  {author} {\bibfnamefont {B.~Y.}\ \bibnamefont {{Sun}}},\ }\href
  {https://doi.org/10.1103/PhysRevC.97.025801} {\bibfield  {journal} {\bibinfo
  {journal} {\prc}\ }\textbf {\bibinfo {volume} {97}},\ \bibinfo {eid} {025801}
  (\bibinfo {year} {2018})},\ \Eprint {https://arxiv.org/abs/1801.08672}
  {arXiv:1801.08672 [nucl-th]} \BibitemShut {NoStop}%
\bibitem [{\citenamefont {{Wan}}\ \emph {et~al.}(2018)\citenamefont {{Wan}},
  \citenamefont {{Xu}}, \citenamefont {{Ren}},\ and\ \citenamefont
  {{Liu}}}]{2018PhRvC..97e1302W}%
  \BibitemOpen
  \bibfield  {author} {\bibinfo {author} {\bibfnamefont {N.}~\bibnamefont
  {{Wan}}}, \bibinfo {author} {\bibfnamefont {C.}~\bibnamefont {{Xu}}},
  \bibinfo {author} {\bibfnamefont {Z.}~\bibnamefont {{Ren}}},\ and\ \bibinfo
  {author} {\bibfnamefont {J.}~\bibnamefont {{Liu}}},\ }\href
  {https://doi.org/10.1103/PhysRevC.97.051302} {\bibfield  {journal} {\bibinfo
  {journal} {\prc}\ }\textbf {\bibinfo {volume} {97}},\ \bibinfo {eid} {051302}
  (\bibinfo {year} {2018})}\BibitemShut {NoStop}%
\bibitem [{\citenamefont {{Lu}}\ \emph {et~al.}(2019)\citenamefont {{Lu}},
  \citenamefont {{Li}}, \citenamefont {{Burgio}}, \citenamefont {{Figura}},\
  and\ \citenamefont {{Schulze}}}]{2019PhRvC.100e4335L}%
  \BibitemOpen
  \bibfield  {author} {\bibinfo {author} {\bibfnamefont {J.-J.}\ \bibnamefont
  {{Lu}}}, \bibinfo {author} {\bibfnamefont {Z.-H.}\ \bibnamefont {{Li}}},
  \bibinfo {author} {\bibfnamefont {G.~F.}\ \bibnamefont {{Burgio}}}, \bibinfo
  {author} {\bibfnamefont {A.}~\bibnamefont {{Figura}}},\ and\ \bibinfo
  {author} {\bibfnamefont {H.~J.}\ \bibnamefont {{Schulze}}},\ }\href
  {https://doi.org/10.1103/PhysRevC.100.054335} {\bibfield  {journal} {\bibinfo
   {journal} {\prc}\ }\textbf {\bibinfo {volume} {100}},\ \bibinfo {eid}
  {054335} (\bibinfo {year} {2019})},\ \Eprint
  {https://arxiv.org/abs/1907.03120} {arXiv:1907.03120 [nucl-th]} \BibitemShut
  {NoStop}%
\bibitem [{\citenamefont {{Steiner}}\ \emph {et~al.}(2013)\citenamefont
  {{Steiner}}, \citenamefont {{Hempel}},\ and\ \citenamefont
  {{Fischer}}}]{2013ApJ...774...17S}%
  \BibitemOpen
  \bibfield  {author} {\bibinfo {author} {\bibfnamefont {A.~W.}\ \bibnamefont
  {{Steiner}}}, \bibinfo {author} {\bibfnamefont {M.}~\bibnamefont
  {{Hempel}}},\ and\ \bibinfo {author} {\bibfnamefont {T.}~\bibnamefont
  {{Fischer}}},\ }\href {https://doi.org/10.1088/0004-637X/774/1/17} {\bibfield
   {journal} {\bibinfo  {journal} {\apj}\ }\textbf {\bibinfo {volume} {774}},\
  \bibinfo {eid} {17} (\bibinfo {year} {2013})},\ \Eprint
  {https://arxiv.org/abs/1207.2184} {arXiv:1207.2184 [astro-ph.SR]}
  \BibitemShut {NoStop}%
\bibitem [{\citenamefont {{Shen}}\ \emph {et~al.}(1998)\citenamefont {{Shen}},
  \citenamefont {{Toki}}, \citenamefont {{Oyamatsu}},\ and\ \citenamefont
  {{Sumiyoshi}}}]{1998PThPh.100.1013S}%
  \BibitemOpen
  \bibfield  {author} {\bibinfo {author} {\bibfnamefont {H.}~\bibnamefont
  {{Shen}}}, \bibinfo {author} {\bibfnamefont {H.}~\bibnamefont {{Toki}}},
  \bibinfo {author} {\bibfnamefont {K.}~\bibnamefont {{Oyamatsu}}},\ and\
  \bibinfo {author} {\bibfnamefont {K.}~\bibnamefont {{Sumiyoshi}}},\ }\href
  {https://doi.org/10.1143/PTP.100.1013} {\bibfield  {journal} {\bibinfo
  {journal} {Progress of Theoretical Physics}\ }\textbf {\bibinfo {volume}
  {100}},\ \bibinfo {pages} {1013} (\bibinfo {year} {1998})},\ \Eprint
  {https://arxiv.org/abs/nucl-th/9806095} {arXiv:nucl-th/9806095 [nucl-th]}
  \BibitemShut {NoStop}%
\bibitem [{\citenamefont {{Shen}}\ \emph {et~al.}(2011)\citenamefont {{Shen}},
  \citenamefont {{Toki}}, \citenamefont {{Oyamatsu}},\ and\ \citenamefont
  {{Sumiyoshi}}}]{2011ApJS..197...20S}%
  \BibitemOpen
  \bibfield  {author} {\bibinfo {author} {\bibfnamefont {H.}~\bibnamefont
  {{Shen}}}, \bibinfo {author} {\bibfnamefont {H.}~\bibnamefont {{Toki}}},
  \bibinfo {author} {\bibfnamefont {K.}~\bibnamefont {{Oyamatsu}}},\ and\
  \bibinfo {author} {\bibfnamefont {K.}~\bibnamefont {{Sumiyoshi}}},\ }\href
  {https://doi.org/10.1088/0067-0049/197/2/20} {\bibfield  {journal} {\bibinfo
  {journal} {\apjs}\ }\textbf {\bibinfo {volume} {197}},\ \bibinfo {eid} {20}
  (\bibinfo {year} {2011})},\ \Eprint {https://arxiv.org/abs/1105.1666}
  {arXiv:1105.1666 [astro-ph.HE]} \BibitemShut {NoStop}%
\bibitem [{\citenamefont {{Oertel}}\ \emph {et~al.}(2017)\citenamefont
  {{Oertel}}, \citenamefont {{Hempel}}, \citenamefont {{Kl{\"a}hn}},\ and\
  \citenamefont {{Typel}}}]{2017RvMP...89a5007O}%
  \BibitemOpen
  \bibfield  {author} {\bibinfo {author} {\bibfnamefont {M.}~\bibnamefont
  {{Oertel}}}, \bibinfo {author} {\bibfnamefont {M.}~\bibnamefont {{Hempel}}},
  \bibinfo {author} {\bibfnamefont {T.}~\bibnamefont {{Kl{\"a}hn}}},\ and\
  \bibinfo {author} {\bibfnamefont {S.}~\bibnamefont {{Typel}}},\ }\href
  {https://doi.org/10.1103/RevModPhys.89.015007} {\bibfield  {journal}
  {\bibinfo  {journal} {Reviews of Modern Physics}\ }\textbf {\bibinfo {volume}
  {89}},\ \bibinfo {eid} {015007} (\bibinfo {year} {2017})},\ \Eprint
  {https://arxiv.org/abs/1610.03361} {arXiv:1610.03361 [astro-ph.HE]}
  \BibitemShut {NoStop}%
\bibitem [{\citenamefont {{Wang}}\ and\ \citenamefont
  {{Chen}}(2017)}]{2017PhLB..773...62W}%
  \BibitemOpen
  \bibfield  {author} {\bibinfo {author} {\bibfnamefont {R.}~\bibnamefont
  {{Wang}}}\ and\ \bibinfo {author} {\bibfnamefont {L.-W.}\ \bibnamefont
  {{Chen}}},\ }\href {https://doi.org/10.1016/j.physletb.2017.08.007}
  {\bibfield  {journal} {\bibinfo  {journal} {Physics Letters B}\ }\textbf
  {\bibinfo {volume} {773}},\ \bibinfo {pages} {62} (\bibinfo {year} {2017})},\
  \Eprint {https://arxiv.org/abs/1705.05122} {arXiv:1705.05122 [nucl-th]}
  \BibitemShut {NoStop}%
\bibitem [{\citenamefont {{Jiang}}\ \emph {et~al.}(2014)\citenamefont
  {{Jiang}}, \citenamefont {{Bao}}, \citenamefont {{Chen}}, \citenamefont
  {{Zhao}},\ and\ \citenamefont {{Arima}}}]{2014PhRvC..90f4303J}%
  \BibitemOpen
  \bibfield  {author} {\bibinfo {author} {\bibfnamefont {H.}~\bibnamefont
  {{Jiang}}}, \bibinfo {author} {\bibfnamefont {M.}~\bibnamefont {{Bao}}},
  \bibinfo {author} {\bibfnamefont {L.-W.}\ \bibnamefont {{Chen}}}, \bibinfo
  {author} {\bibfnamefont {Y.~M.}\ \bibnamefont {{Zhao}}},\ and\ \bibinfo
  {author} {\bibfnamefont {A.}~\bibnamefont {{Arima}}},\ }\href
  {https://doi.org/10.1103/PhysRevC.90.064303} {\bibfield  {journal} {\bibinfo
  {journal} {\prc}\ }\textbf {\bibinfo {volume} {90}},\ \bibinfo {eid} {064303}
  (\bibinfo {year} {2014})}\BibitemShut {NoStop}%
\bibitem [{\citenamefont {{Xu}}\ \emph {et~al.}(2007)\citenamefont {{Xu}},
  \citenamefont {{Chen}}, \citenamefont {{Li}},\ and\ \citenamefont
  {{Ma}}}]{2007PhRvC..75a4607X}%
  \BibitemOpen
  \bibfield  {author} {\bibinfo {author} {\bibfnamefont {J.}~\bibnamefont
  {{Xu}}}, \bibinfo {author} {\bibfnamefont {L.-W.}\ \bibnamefont {{Chen}}},
  \bibinfo {author} {\bibfnamefont {B.-A.}\ \bibnamefont {{Li}}},\ and\
  \bibinfo {author} {\bibfnamefont {H.-R.}\ \bibnamefont {{Ma}}},\ }\href
  {https://doi.org/10.1103/PhysRevC.75.014607} {\bibfield  {journal} {\bibinfo
  {journal} {\prc}\ }\textbf {\bibinfo {volume} {75}},\ \bibinfo {eid} {014607}
  (\bibinfo {year} {2007})},\ \Eprint {https://arxiv.org/abs/nucl-th/0609035}
  {arXiv:nucl-th/0609035 [nucl-th]} \BibitemShut {NoStop}%
\bibitem [{\citenamefont {{Agrawal}}\ \emph {et~al.}(2014)\citenamefont
  {{Agrawal}}, \citenamefont {{De}}, \citenamefont {{Samaddar}}, \citenamefont
  {{Centelles}},\ and\ \citenamefont {{Vi{\~n}as}}}]{2014EPJA...50...19A}%
  \BibitemOpen
  \bibfield  {author} {\bibinfo {author} {\bibfnamefont {B.~K.}\ \bibnamefont
  {{Agrawal}}}, \bibinfo {author} {\bibfnamefont {J.~N.}\ \bibnamefont {{De}}},
  \bibinfo {author} {\bibfnamefont {S.~K.}\ \bibnamefont {{Samaddar}}},
  \bibinfo {author} {\bibfnamefont {M.}~\bibnamefont {{Centelles}}},\ and\
  \bibinfo {author} {\bibfnamefont {X.}~\bibnamefont {{Vi{\~n}as}}},\ }\href
  {https://doi.org/10.1140/epja/i2014-14019-8} {\bibfield  {journal} {\bibinfo
  {journal} {European Physical Journal A}\ }\textbf {\bibinfo {volume} {50}},\
  \bibinfo {eid} {19} (\bibinfo {year} {2014})},\ \Eprint
  {https://arxiv.org/abs/1308.5527} {arXiv:1308.5527 [nucl-th]} \BibitemShut
  {NoStop}%
\end{thebibliography}%

\end{document}